\documentclass[useAMS,usenatbib]{mn2e}

\usepackage[psamsfonts]{amssymb}
\usepackage{graphicx} 
\usepackage{amsmath}
\usepackage{color}
\usepackage{multirow}
\usepackage{rotating}
\usepackage[draft]{hyperref}
\usepackage{pdflscape}
\usepackage{supertabular}
\usepackage[english]{babel}
\usepackage{bbm}
\usepackage{booktabs}  
\DeclareMathOperator\erf{erf}		

\newcommand{\Msol}{\;\mathrm{M}_{\odot}}

\begin{document}
                                                                   
\makeatletter		
\def\fps@figure{htbp}
\makeatother                                                                   
                                                                                                          
\title[When do star clusters become multiple star systems? II]{When do star clusters become multiple star systems? II.  
Toward a half-life formalism with four bodies}
\author[Ibragimov, Leigh, Ryu, Panurach \& Perna]{Timur Ibragimov$^{1}$, Nathan W. C. Leigh$^{1,2,3}$\thanks{E-mail: nleigh@amnh.org (NWCL)}, Taeho Ryu$^{2}$, Teresa Panurach$^{1,4}$, \newauthor{Rosalba Perna$^{2}$}\\
$^{1}$Department of Astrophysics, American Museum of Natural History, Central Park West and 79th Street, New York, NY 10024, USA\\
$^{2}$Department of Physics and Astronomy, Stony Brook University, Stony Brook, NY 11794-3800, USA\\
$^{3}$Center for Computational Astrophysics, Flatiron Institute, 162 Fifth Avenue, New York, NY 10010, USA\\
$^{4}$Department of Physics and Astronomy, Hunter College, City University of New York, New York, NY, 10065, USA}

\pagerange{\pageref{firstpage}--\pageref{lastpage}} \pubyear{2017}

\maketitle

\label{firstpage}

\begin{abstract}
We present a half-life formalism for describing the disruption of gravitationally-bound few-body systems, with a focus on binary-binary scattering.  For negative total encounter energies, the four-body problem has three possible decay products in the point particle limit.  For each decay product and a given set of initial conditions, we obtain directly from numerical scattering simulations the half-life for the distribution of disruption times.  As in radioactive decay, the half-lives should provide a direct prediction for the relative fractions of each decay product.  We test this prediction with simulated data and find good agreement with our hypothesis.  We briefly discuss applications of this feature of the gravitational four-body problem to populations of black holes in globular clusters. This paper, the second in the series, builds on extending the remarkable similarity between gravitational chaos at the macroscopic scale and radioactive decay at the microscopic scale to larger-$N$ systems.
\end{abstract}


\section{Introduction} \label{intro}

The disruption of gravitationally-bound systems of chaotically-interacting particles is of direct relevance to a number of sub-fields within astrophysics.  For example, the disruption of open star clusters a few tens of millions of years after they are born is the underlying mechanism populating the Galactic disk.  Similarly, several authors have argued that the Galactic bulge formed from the disruption of globular clusters and/or minor mergers with satellite galaxies, stripped of their stars by the tidal field of the inner Galaxy to the point of disintegration \citep[e.g.][]{binney87,webb14,webb15}.  The disruption of bound systems of stars hosting a super-massive black hole can produce hypervelocity stars, some of which are observed in the halo of the Milky Way \citep[e.g.][]{brown05,brown12}. 

A particularly timely example of the importance of the disruption of few-body systems in astrophysics pertains to populations of stellar-mass black holes (BHs) in globular clusters (GCs) \citep[e.g.][]{askar17,rod17}, which are often argued to contribute non-negligibly to the rate of binary BH mergers detected by aLIGO. For instance, dynamically-formed BH-BH mergers yield up to 100 coalescence events per year per Giga-parsec in GCs \citep{rod15}.
BHs tend to be the most massive objects in GCs, such that energy equipartition with the background stellar field reduces the mean BH velocities, causing them to sink to the cluster core.  These mass segregated BHs then form a sub-cluster in the core.  Depending on the masses of the BHs and the total masses of the BH and stellar systems in the core, one of two fates are possible for such a BH sub-cluster \citep{spitzer87}.  If the mass of the BHs does not dominate the local potential, then the BH sub-cluster remains in steady-state.  If, however, the mass of the BH population dominates over the local stellar mass, then an instability can set in, in which case the BH sub-cluster contracts \citep{spitzer69}.  Subsequently, its density continues to rise until individual BHs (and BH binaries) begin to experience strong encounters with other BHs and BH binaries.
The net effect of these strong interactions is the continued ejection of BHs from the core, one by one, and their escape from the entire GC.  This process ultimately continues until only a few BHs remain \citep[e.g.][]{kulkarni93,sigurdsson93,leigh14}.  The end state of the BH population in this scenario is a completely disrupted sub-cluster, with most BHs ejected outside of the GC to join the Galactic field, most likely in the halo.  \footnote{More recent studies have shown that this generic result is quite sensitive to the cluster mass and concentration \citep[e.g.][]{breen13a,breen13b}.  In particular, the rate of evolution of the BH sub-system is determined by the rate of energy flow across the half-mass radius, instead of the local relaxation time.  This amendment to the original formalism predicts that many GCs should host non-negligible numbers of BHs, since the time-scale for their ejection from the cluster can exceed a Hubble time for many Galactic GCs \citep[e.g.][]{breen13a,breen13b}.} 

In Paper I of this series \citep{leigh16b}, we studied the distributions of disruption times for chaotic gravitational encounters as a function of the number of interacting particles.  We calculated the distributions of disruption times as a function of both the particle number $N$ and the virial coefficient (defined as the ratio of the total system kinetic energy to the total potential energy).  The subsequent distributions were fit with a physically-motivated function, consisting of an initial exponential decay followed by a very slowly decreasing tail at long encounter times.  We attributed these tails to either long-lived excursions of individual particles or prolonged quasi-stable interactions (or a combination of both), based on similar studies of the three-body problem, but left the exact physical mechanism responsible as an open question.  

We found three primary features characteristic of the calculated distributions of disruption times:  (1) the system half-life increases with increasing particle number, (2) the fraction of long-lived (excursive? quasi-stable? -- the mechanism underlying the power-law tail at longer encounter durations remains poorly understood) encounters increases with increasing particle number and (3) both the system half-life and the fraction of long-lived encounters increase with decreasing virial coefficient.  This last trend is likely due to a deviation from resonant encounters, and hence the assumption of ergodicity.

In this paper, the second in the series, we continue our study of the distributions of disruption times for point- particles in chaotic Newtonian gravity, for different combinations of particle masses, and use it to derive a half-life formalism for describing the disruption of gravitationally-bound N-body systems. Throughout this study, we will use $N =$ 4 as a sample application of our formalism. This sample size is favorable since the computational expense of increasing $N$ in simulations is scaled by a factor of $N^2$ \citep[e.g.][]{dehnen11}.  Hence, we are able to maximize the number of simulations performed for maximal statistical significance of our results. 

In \textsection~\ref{model}, we present a combinatorics-based formalism for calculating outcome branching ratios using half-lives, assuming point-particles.  In \textsection~\ref{method}, we describe the simulations used in this study to test our theoretical model, as well as the approaches used to obtain our results.  We derive the simulated half-lives and use them to calculate branching ratios in \textsection~\ref{results}, which we subsequently compare directly to the simulated branching ratios. In order for our half-life formalism to work, this constrains the time between successive ejection events during outcomes in which two single stars escape.  To test this and confirm the validity of our model, we perform additional scattering experiments in \textsection~\ref{confirmation}. The assumptions underlying our model and their applicability to astrophysical systems are discussed in \textsection~\ref{discussion}, along with future work.  Our key results are summarized in \textsection~\ref{summary}.

 
\section{Model} \label{model}

In this section, we present a combinatorics-based formalism for calculating outcome branching ratios using half-lives.  We assume identical (and, briefly, distinct) point-particles throughout this section. This means that we group all mass combinations together in the final outcome, as if the particle masses are unknown.  As will be demonstrated in the next section, this limits the number of final outcome states we consider to three (i.e., $2+1+1$, $2+2$ and $3+1$), and ensures that the total number of simulations resulting in each outcome is sufficiently large that the final lifetime distributions have converged. Later, we relax this assumption, to consider the lifetime distributions of decay products with different mass combinations.

\subsection{Combinatorial half-life formalism} \label{virial}

In this section, we present and justify the number of distinct decay products of the four-body problem involving point particles (and negative total encounter energies) considered throughout this paper.  First, let us consider the case of distinguishable particles (e.g., if the particle masses are different and can be measured in the final outcome state).  For the decay of $N$ distinct particles, the number of possible hierarchical arrangements of the particles in the final configuration is:
\begin{equation}
\label{eqn:comb1}
N_{\rm h} = \sum_{i=0}^{N-2} {N - i\choose 2}.
\end{equation}  

\noindent For $N =$ 4, we have $N_{\rm h} =$ 10.\\
For our purposes, this is a large number of outcomes to consider.  Specifically, for a given total number of simulations, the sample size for a given outcome and the statistical significance of the results decreases as the number of outcomes increases. Consequently, it is in our interest to consider as few outcomes as possible.

Next, let us consider the case that all particles are indistinguishable (e.g., the particle masses are different, but it is not possible to know which particle has a given mass in the final state).  Here, the total number of possible state outcomes $N_{\rm s}$ is determined by an appropriate partition function:
\begin{equation}
\label{eqn:comb2}
N_{\rm s} = \sum_{i=1}^{N-1} P_{i}(N, k).
\end{equation} 
Equation~\ref{eqn:comb2} has no simple analytic solution, and must be evaluated on a case-by-case basis.  Equation~\ref{eqn:comb2} describes the number of generic outcome states for the binary-binary scattering simulations presented in this study.  For $N =~$4, we have $N_{\rm s} = $ 3 for total encounter energies $E \le$ 0.  That is, when an encounter is over, the remaining configuration is described by one of the following outcomes \citep{leigh16b}:
{\renewcommand{\labelitemi}{$\triangleright$}
\begin{itemize}
\item two binaries (2+2; $k =$ A)
\item a triple and a single star (3+1; $k =$ B)
\item a binary and two runaway singles (2+1+1; $k =$ C)
\end{itemize}}
Although the complete ionization or the 1+1+1+1 outcome, is possible, it is, in practice, an exceedingly rare occurrence for $E\approx~$0 \citep{ryu17a}. 

As a result of this simple exercise, we consider all particles to be indistinguishable throughout this paper, independent of  mass distribution in the final state. This minimizes the total number of generic encounter outcomes that we must consider, and hence maximizes the statistical significance of the simulated lifetime distribution for each outcome.

For the chaotic four-body problem involving indistinguishable particles, we assume that each of our three final outcome states has an associated half-life $\tau_{k}$ (where $k$ runs from A to C; see above). 
Then, if we begin with $N_{\rm 0}$ initial on-going (i.e., resonant) four-body interactions, the total number of expected decay products $N_{\rm d}$ after a total time $t$ is:
\begin{equation}
\label{eqn:comb3}
N_{\rm d}(t) = N_{\rm 0}(1-e^{\rm -{\lambda}t}),
\end{equation}  
where
\begin{equation}
\label{eqn:comb4}
\lambda = \sum\limits_{ k={\rm A,B,C}} {\lambda_{\rm k}},
\end{equation} 
and $N_{0}$ is the initial number of $4$-body systems that are to decay. In this paper, this corresponds to 40,000 simulated experiments, thus here $N_{0}$ was 40,000, large enough to satisfy the large sample theory (physically prevalent to unstable star clusters like GCs) and corroborate Equation~\ref{eqn:comb3}.

Important to note is the close resemblance of Equation \ref{eqn:comb3} with the general first-order differential solution characteristic of models of radioactive decay. The parameter $\lambda_{k}$ is the decay constant of the $k$ outcome, and is associated with the system half-life via the relations:
\begin{equation}
\label{eqn:comb5}
\lambda_{k} = \frac{1}{\tau_{k}}
\end{equation} 
and
\begin{equation}
\label{eqn:comb6}
\tau_{k,1/2} = \frac{\ln{2}}{\lambda_{ k}} = \tau_{k}\ln{2}.
\end{equation} 

The half-life associated with a given outcome state $\tau_{k,1/2}$ can be found directly from fitting a hyperbolic-trigonometric (i.e., exponential) function to the lifetime distribution found from numerical scattering experiments, as described in Paper I of this series \citep{leigh16b}. Alternatively, a proxy for the half-life can be found using Gaussian fits to the simulated lifetime distributions due to their similarly exponential functional form, as will be shown later in this enterprise.

We note that, assuming indistinguishable particles implies that different combinations of particle masses cannot be distinguished in the decay products.  That is, it is not possible to measure the particle masses in either the initial or final states.  The particle masses are unknown.  Consequently, in \textsection~\ref{fittingmethod}, we group together all mass combinations within the same outcome.  We defer an exploration of the half-life dependences for different combinations of particle masses in each decay product to \textsection~\ref{masscombs}.

\subsection{Branching ratios} \label{branching}

Given the half-lives $\tau_{k,1/2}$ of all outcome states associated with a given number of (indistinguishable) particles $N$, we can calculate the fraction of numerical scattering simulations expected to yield each final state:
\begin{equation}
\label{eqn:comb7}
f_{k} = \frac{\lambda_{k}(t)}{\lambda(t)},
\end{equation} 
for any time $t$, where:
\begin{equation}
\label{eqn:comb8}
1 = \sum_{k=1}^{\rm N_{\rm s}} f_{k}.
\end{equation}

\section{Method} \label{method}

In this section, we present the numerical scattering experiments and methods of data analysis used to study binary-binary encounters involving point-particles, including different combinations of particle masses.  Our goal is to present and justify our methods of data analysis, including the primary $N$-body integrator \texttt{FEWBODY} used in this paper, along with a description of the scattering trials performed.  Ultimately, we wish to test the hypothesis that the outcomes of the four-body problem involving point particles can be described using a half-life formalism, in analogy with radioactive decay.

\subsection{Numerical scattering experiments} \label{exp}

We calculate the outcomes of a series of binary-binary (2+2) encounters using the \texttt{FEWBODY} numerical 
scattering code\footnote{For the source code, see http://fewbody.sourceforge.net.}.  The code integrates the usual 
$N$-body equations in configuration- (i.e., position-) space in order to advance the system forward in time, using the eighth-order Runge-Kutta Prince-Dormand integration method with ninth-order error estimate and adaptive time-step.  
For more details about the \texttt{FEWBODY} code, we refer the reader to \citet{fregeau04}.  

We perform three different sets of fiducial simulations, as in \citet{leigh16b}.  The first set assumes particle masses of $m_{\rm 1} =10 \Msol$ and $m_{\rm 2} = 1 \Msol$, distributed randomly among the initial particles.  The other sets of simulations are analogous to the first, but here we adopt particle masses of either $m_{\rm 1} = 5 \Msol$, or $m_{\rm 1} = 3 \Msol$. 

In all simulations all binaries have $a_{\rm 1} =$ $a_{\rm 2} =$ 5 AU initially, and
eccentricities $e_{\rm 1} = e_{\rm 2} =$ 0---thus circular orbits of radius 5 AU. We set the impact parameter to zero and the initial relative velocity at infinity $v_{\rm rel}$ to 0.5$v_{\rm crit}$, where $v_{\rm crit}$ is the critical velocity and is defined as the relative
velocity at infinity needed for a total encounter energy of zero.\footnote{Note that this choice of relative velocity is somewhat arbitrary, but is typical for dense star clusters \citep[e.g.][]{geller15,leigh16a}.}
The angles defining the initial relative configurations of the binary orbital planes and phases are chosen at random.  These initial conditions ensure that the total encounter energies span a relatively narrow range across every set of simulations.  However, this approximation breaks down more as the mass ratio deviates from unity. We we will return to this issue in \textsection~\ref{discussion}. Though these initial conditions do not accurately portray observed star-cluster environments, given the theoretical nature of this study, these starting simplifications are necessary. 

We perform 4 $\cdot$ 10$^4$ numerical scattering experiments for every combination of particle masses in the $10\Msol/1\Msol$ system, resulting in a combined 2 $\cdot$ 10$^5$ simulations. The same amount of scattering experiments are performed for the $3\Msol/1\Msol$ and $5\Msol/1\Msol$ systems, yielding a net $6 \cdot 10^5$ simulations.

All simulations are terminated using the same criteria as \citet{fregeau04} to decide 
when a given encounter is complete, and we refer the reader to that paper for more details. We also impose a maximum integration time of 10$^6$ years (i.e., if the particles are still interacting chaotically at this time, then the simulations terminate artificially), and emphasize that one final orbital period of the resultant must elapse before the integration ceases (we will return to this in Section~\ref{results}).  Finally, each hierarchy must also be dynamically 
stable and experience a tidal perturbation from other nodes within the same hierarchy that is less than the critical value 
adopted by \texttt{FEWBODY}, called the tidal tolerance parameter.  For this study, we adopt a tidal tolerance parameter 
$\delta =$ 10$^{-7}$ for all simulations.  This choice for $\delta$ is arbitrary but is sufficiently strict to ensure the accuracy of our simulations 
in the point-particle limit while also balancing the computational expense, particularly at low virial ratios (see \citealt{geller15} and \citealt{leigh16a} for more details).  In particular, larger tidal tolerance parameters can result in spurious or unphysical triple formation, which we avoid with this choice for $\delta$.

\begin{figure}
\centering
\includegraphics[width=\columnwidth]{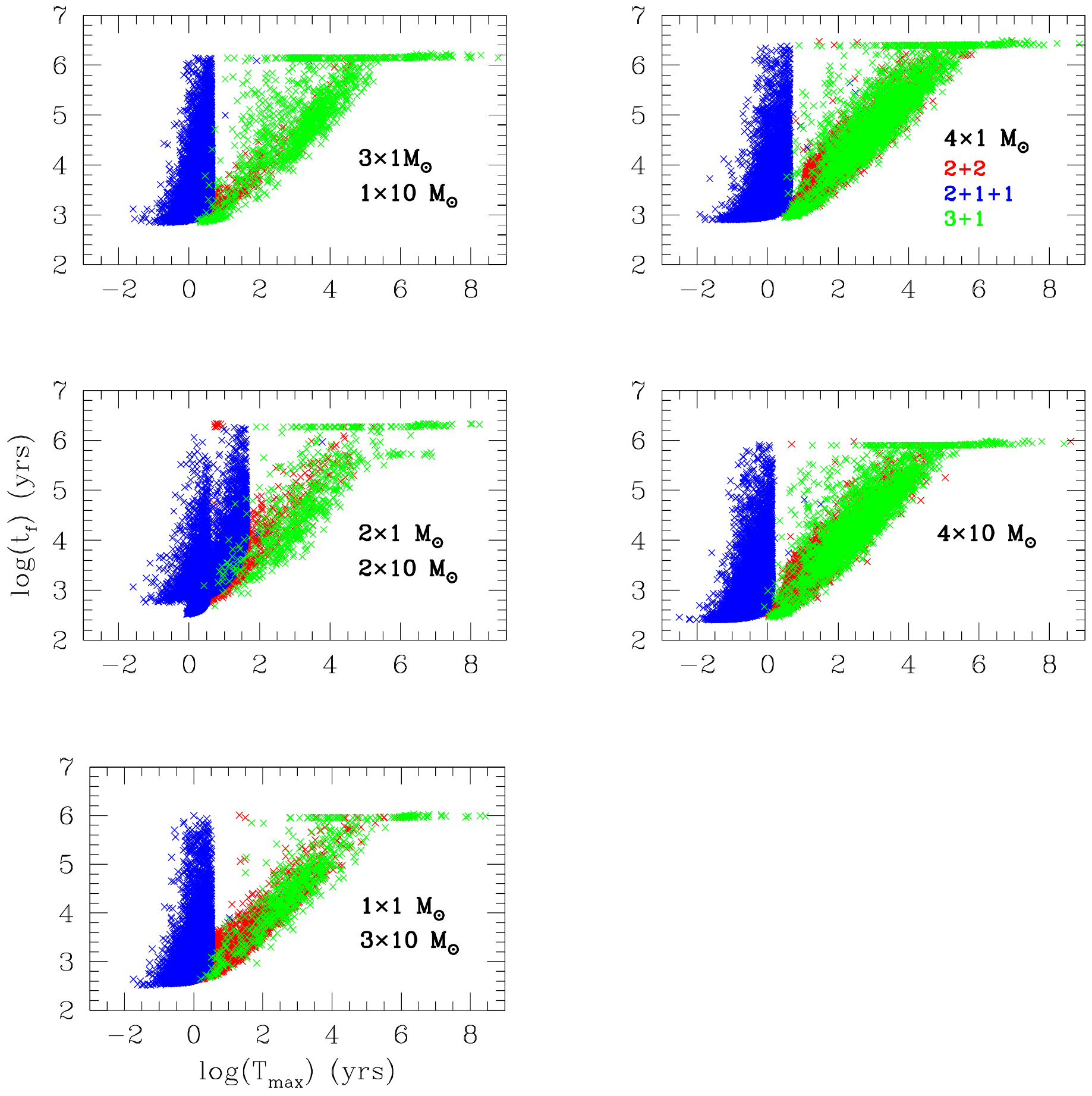}
\caption[Relation between the maximum orbital period and the lifetime for mass ratio $m_{\rm 1}/m_{\rm 2} = 10$.]{Relation between the maximum final orbital period $T_{\rm max}$ of each decay product and the lifetime $t_{\rm f}$ for the mass combination $m_{1} = 10 \Msol$ and $m_2 = 1 \Msol$. Note that $\tau_{0}$ is a normalization factor to make the argument in the logarithm dimensionless. We take $\tau_{0}=1$ yr. The resultants are color schemed and the key is included in the top right inset. Notice the outlier patches gathered in a horizontal arrangement at the top-right corner of each plot. The strength of these horizontal patches corresponds directly to the amount of hypothesized extraneousness presented by the simulation software, and the need to remove it. In addition, the green ($3+1$) and red ($2+2$) resultants have a completely different behavior than the blue in terms of regression. This similarity will be displayed in other visual representations, such as Figures \ref{fig:10CHistograms} and analogs.}
\label{fig:10Correlations}
\end{figure}

\begin{figure}
\centering
\includegraphics[width=\columnwidth]{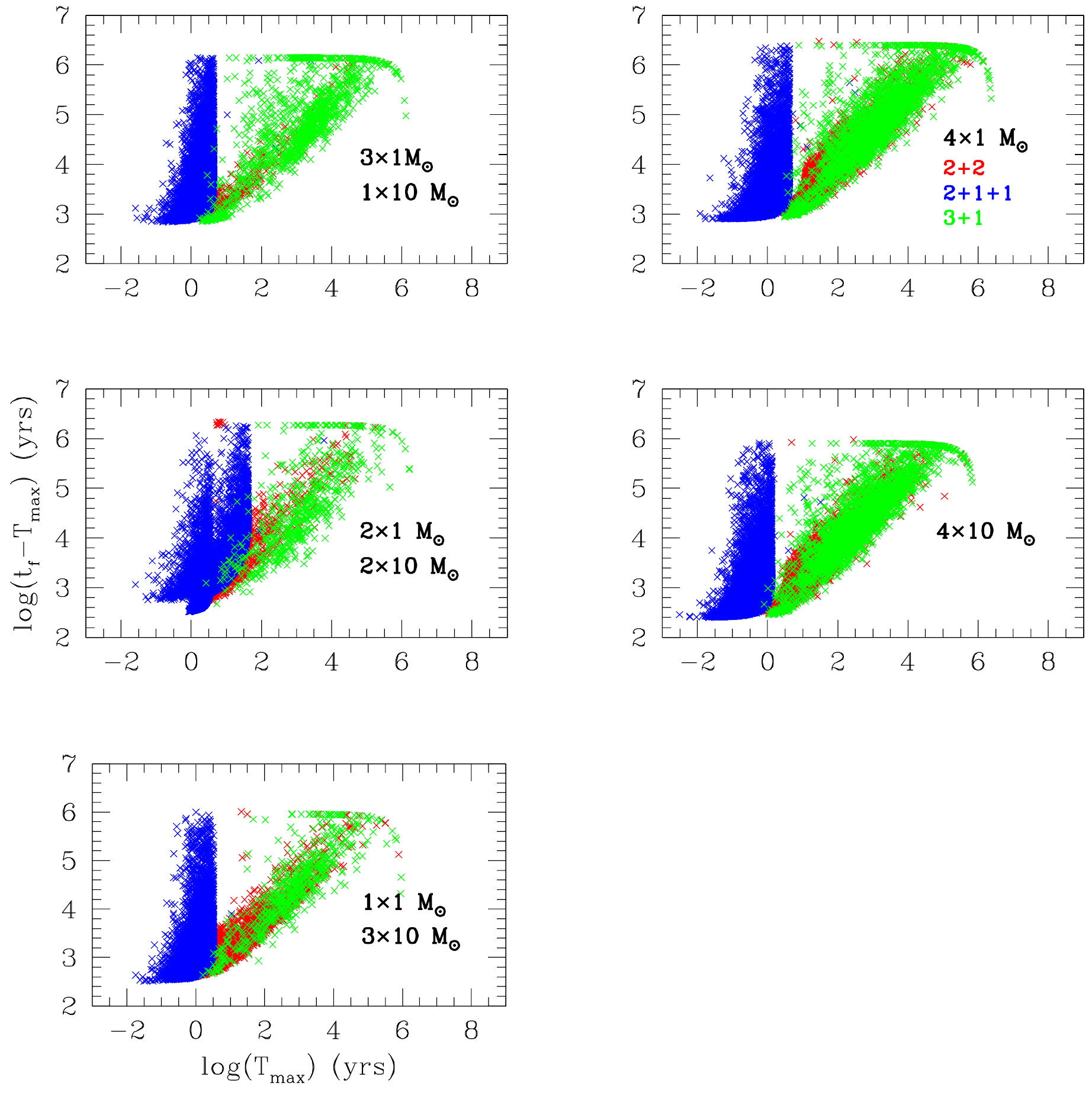}
\caption[Relation between the maximum orbital period and the total simulation lifetime for mass ratio $m_{\rm 1}/m_{\rm 2} = 10$, illustrating the need to subtract $T_{\rm max}$.]{Analogous to Figure \ref{fig:10Correlations} for the mass ratio $m_{\rm 1}/m_{\rm 2} = 10$, but after subtracting $T_{\rm max}$ from the total simulation lifetime $t_{\rm f}$.  This rids us of an artificial pile-up of simulations at $\log (t_{\rm f}/{\rm yr}) = 6$, which is our chosen maximum integration time for the simulations.  If a given interaction is still on-going and in a chaotic state at this time, the simulation is terminated artificially. As a result, throughout the paper, our corrected disruption time used for all histograms is the absolute difference between $T_{\rm max}$ and $t_{\rm f}$.
This subtraction is a great asset to the visual representations, as seen in Figure \ref{fig:10Models} and analogs, and is also evident in those graphs by a significantly diminished long-period bin.}
\label{fig:10SubtCorrelations}
\end{figure}

Though the aforementioned threshold parameters are necessary, they may influence the time-dependent outcome of interacting bodies, namely at long-excursions.  This is because \texttt{FEWBODY} completes one orbital cycle in the final decay products before outputting the disruption time \citep{leigh16b}.  
This, combined with the long-period tail following the initial exponential drop of the cumulative lifetime distribution, prompted us to subtract  $T_{\rm max}$ from the disruption time $t_{\rm f}$ for the time-axis of all graphs, where $T_{\rm max}$ represents the longest orbital period of each final decay product. We suspect that this phenomenon is indicated by the horizontal congregation of points at the top right of each plot in Figure~\ref{fig:10Correlations}---a non-physical extraneous occurrence that is due to \texttt{FEWBODY}'s method of considering an integration complete---thus attaining $t_{\rm f}$---only when one final period of each remaining hierarchy has elapsed. The subtraction of $T_{\rm max}$ allows us to consider the more accurate disruption time, which we define as $\tau_{\rm d} = t_{\rm f} - T_{\rm max}$. For almost all simulated instances,  $t_{\rm f} \gg T_{\rm max}$.  This is even true for most $3+1$ outcomes where a stable inner binary is orbited by a long-period \lq\lq secondary."  Thus, this subtraction negligibly affects most values of $\tau_{\rm d}$, save for several outliers.
Stated another way, our goal in this study is to compare the times of disruption for the different outcomes of the four-body problem.  We define the disruption time as the time at which point stability is first achieved in the final decay products, to facilitate a consistent comparison between the different encounter outcomes.  Since \texttt{FEWBODY} integrates over the final orbital period of the widest orbit in each decay product, this disruption time is formally $\tau_{\rm d} = t_{\rm f} - T_{\rm max}$.  This means that the values on the y-axis can become negative in Figure~\ref{fig:10SubtCorrelations}, but we do not include these few negative interaction times in our final distributions. This procedure is necessary to eliminate the integrator artifacts which may misrepresent their physical counterparts. In particular, whenever the simulation is terminated at 10$^6$ years, on the rare occasions in which final orbital period is longer than this and the integrator closes the final orbit, it is not possible to ascertain when stability was first achieved.

As seen in Figure~\ref{fig:10SubtCorrelations}, the above procedure results in a significant reduction of the horizontal long-period line at the top right of each inset. Consequently, all the $y$-axes explicitly mentioning $\tau_{\rm d}$ in this paper actually represent the corrected disruption time with subtracted $T_{\rm max}$ from $t_{\rm f}$.\\
\indent Notice the remaining bin at $\log(\tau_{\rm d}/{\rm yr})\approx6$ years even
after the subtraction of $T_{\rm max}$, which corresponds to the remaining horizontal batch of points in Figure~\ref{fig:10SubtCorrelations}. Being a likely artifact of still on-going interactions at the critical cut-off time, this long-period bin did not affect our manual fitting procedures, since the fraction of the total number of simulations in this bin is negligible (i.e., roughly a few percent of the total number of simulations performed) and we focused on the data near the center of the distribution curves when performing our fits.

It is interesting to note that the lifetime distributions for the $2+1+1$ outcome shows a relatively steep cut-off along the $T_{\rm max}/\tau_0$ axis. Physically, this arises because the two single stars escape with positive total energy, such that the maximum orbital energy of the final binary is equal to the total encounter energy (i.e., in the limit that both single stars escape with velocities that asymptote to zero at large distances and long times). For the simulation sets with different mass combinations, the position of the cut-off varies due to different total energies. In addition, this cut-off is pronounced mainly due to a logarithmic-scaling of the horizontal axis, which would become less sharp and more diffuse with regular graphical representations of disruption time.

\section{Results} \label{results}

In this section, we present the results of our numerical scattering simulations and obtain half-lives for the lifetime distributions.  We focus on the mass ratio $m_{\rm 1}/m_{\rm 2} = 10$ in this section, and present the results for all other mass ratios in an Appendix.

\subsection{Fitting the lifetime distributions}
\label{fittingmethod}

\begin{figure}
\centering\includegraphics[width=\columnwidth]{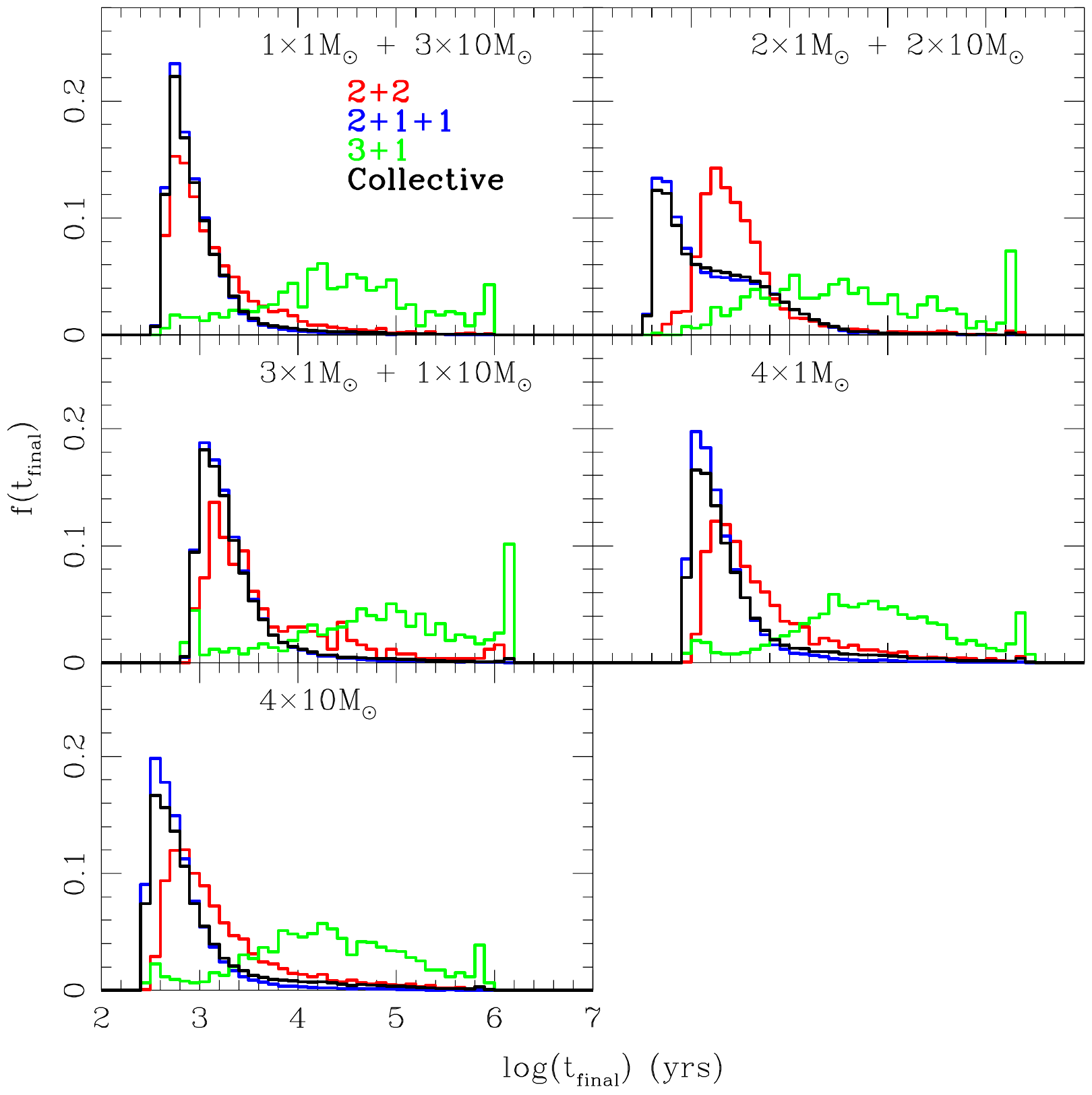}
\caption[Combined distributions and partitioned into resultants for the mass ratio $m_{\rm 1}/m_{\rm 2} = 10$]{The lifetime distributions for all combined simulations, along with the conditional partitioning into the three resultants, or decay products, assuming $m_{\rm 1} =10\Msol$ and $m_{\rm 2} =1 \Msol$.  The different panels show the results for different combinations of the numbers of each particle type, as indicated in the upper right inset of each panel.  The green, blue and red curves show the distributions for, respectively, the 3+1, 2+1+1 and 2+2 outcome states. Notice the strong Gaussian resemblance of these histograms, especially for the $3+1$ case. Furthermore, notice how the black (collective) histogram that represents the total lifetime distribution almost neatly aligns with the blue histograms, indicating that the $2+1+1$ outcome is most likely to occur over the \emph{entire interval} of the simulations.  In other words, most of our binary-binary simulations result in the $2+1+1$ outcome ($>$80\%), such that showing the total (i.e. collective) lifetime distribution (black) resembles most closely the lifetime distribution for the $2+1+1$ outcome.}
\label{fig:10CHistograms}
\end{figure}

As shown in Figures~\ref{fig:10CHistograms}, \ref{fig:3CHistograms}, and \ref{fig:5CHistograms}, the $2+1+1$  is the resultant that determines how the combined (i.e., collective; black lines) histograms will appear. This makes sense, considering that $N_{2+1+1}\gg N_{2+2}\gtrapprox N_{3+1}$, as shown in the plots of relative branching ratios---Figures~\ref{fig:10Fractions}, \ref{fig:3Fractions}, and \ref{fig:5Fractions}.  It follows that the $2+1+1$ outcome's interval of dominance is what determines the overall normalized lifetime distribution.  In other words, most of our binary-binary simulations result in the $2+1+1$ outcome ($>$80\%; see Figure~\ref{fig:10Fractions}), such that showing the total (i.e. collective) normalized lifetime distribution (black) resembles most closely the lifetime distribution for the $2+1+1$ outcome.  It follows that the total half-life, as given by Equation~\ref{eqn:comb6}, should be roughly equal to the half-life for the $2+1+1$ outcome.

Each of the three resultants in Figures \ref{fig:10CHistograms}, \ref{fig:3CHistograms}, and \ref{fig:5CHistograms} resembles a Gaussian curve with various widths or standard deviations, amplitudes, and horizontal shifts.  For modeling purposes, each of these were accounted for by a free-parameter, totaling three: $1/C$ for amplitude-correction so the curves integrate to unity, $t_0$ for a rightward shift,
and $\sigma$ for widths (and, in a physical context, the gauge for half-life $\tau_{k, 1/2}$). Thus, in an attempt to mathematically model the fractions received from the histograms, we manually (i.e., by eye) superimposed Gaussian models of fit of the following form\footnote{Here, $\tau_{0}$ is a normalization factor to make the argument in the logarithm dimensionless by having it as a divisor. In this paper, $\tau_{0}=1$ yr.  We note that this choice for $\tau_{0}$ is arbitrary, and does not affect the relevant parameters obtained from our fitting functions.}:\\
\begin{equation}
f_k(\log(\tau_d/\tau_{0}))=\frac{1}{C}\exp\Big[-\frac{1}{2}\left(\frac{\log(\tau_{\rm d}/t_{\rm k,0})}{\sigma/\tau_{0}}\right)^{2}\Big],
\label{eqn:Gaussian}
\end{equation}
and:
\begin{align}
C& =\int_0^\infty \exp\left[-\frac{1}{2}\left(\frac{\log(\tau'_{\rm d}/\tau_0)-\log(t_0/\tau_0)}{\sigma/\tau_{0}}\right)^2\right]~d(\log(\tau'_{\rm d}/\tau_0))\nonumber\\
& =  \frac{|\sigma|}{\tau_{0}}\sqrt{\frac{\pi}{2}}~\erf\left[\frac{\log(\tau'_{\rm d}/\tau_0)-\log(t_{0}/\tau_0)}{\sqrt{2}~|\sigma|/\tau_{0}}\right]\Bigg|_{\tau'_{\rm d}=0}^{\tau'_{\rm d}\to\infty}\nonumber\\
& = \sqrt{2\pi}\cdot\frac{\sigma}{\tau_{0}}
\end{align}
From this, it follows that $f_k(\log(\tau_{\rm d}/\tau_{0}))$ is a \emph{probability density function} (\emph{PDF}) because the area under the $f_k$ curve from $0$ to $\infty$ is virtually equal to the area from $-\infty$ to $\infty$, which equals 1 (under a normalized $y$-axis for fraction) \citep{calculus98}.

We emphasize that a Gaussian accurately describes the $3+1$ resultant over the entire range, but only the Primary half of a Gaussian is needed for each of the $2+1+1$ and $2+2$ outcomes.  Due to these inconsistencies, and the need for a Secondary Gaussian to fully fit the $2+1+1$ and $2+2$ lifetime distributions (see below), we performed these fits by eye.  This means that we fit the portion of the lifetime distributions closest to the peak with a \lq\lq Primary" Gaussian, and then use a \lq\lq Secondary" Gaussian to fit the remaining part of the distribution.  Due to this approach, we do not obtain formal best-fit parameters for each Gaussian, and instead maximize the goodness-of-fit over the interval of interest.   
This fitting procedure is more than sufficient for the goal of this project in developing a general framework for disruption distributions.

\begin{figure}
\centering\includegraphics[width=\columnwidth]{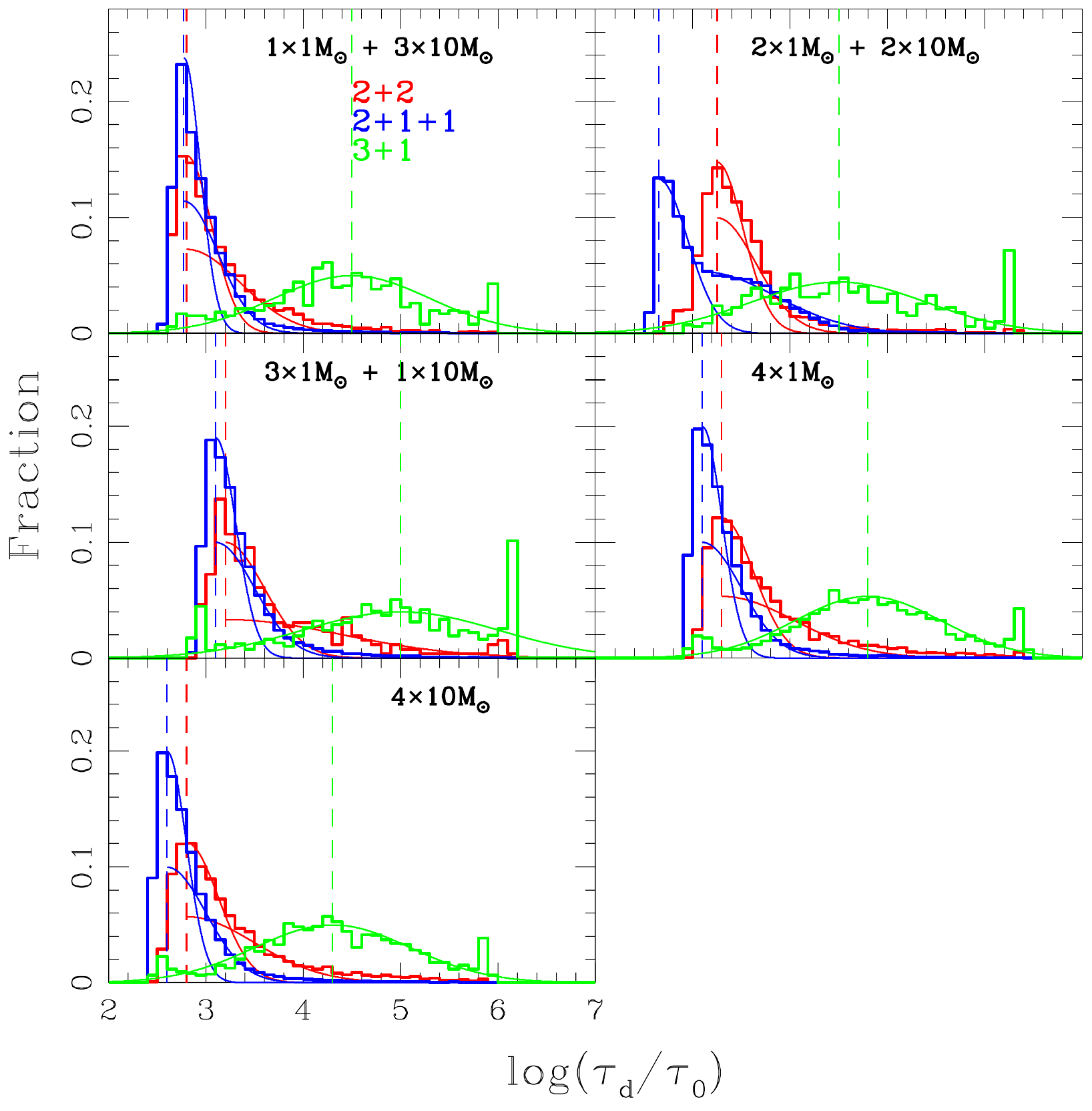}
\caption[The distributions of encounter durations for each possible outcome state, and for all combinations of light and heavy particles, with $m_{\rm 1} = 10 \Msol$]{The same as Figure~\ref{fig:10CHistograms}, but superimposed are the models of fit using Equation~\ref{eqn:Gaussian}, manually created and color-coded to their corresponding resultant. Only the right halves of the red and blue curves are plotted to demonstrate the interval of interest for these histograms (which we use to develop our half-life formalism). Note that to obtain the half-lives we focus only on the Primary Gaussians, since they fit a majority of the data. In all cases, the Primary Gaussians are those with larger amplitude than the Secondary Gaussians.}
\label{fig:10Models}
\end{figure}

Unlike the $3+1$ histograms in Figure \ref{fig:10Models}, which require only one Gaussian to fully fit the lifetime distributions, the $2+2$ and $2+1+1$ counterparts require two Gaussian curves for each, one Primary for fitting the bulk of the lifetime distribution and one Secondary to include the excess at long encounter durations. The values of all of the fitting constants used to create these plots with Equation~\ref{eqn:Gaussian} are included in Table~\ref{table:table}.  One indirect method to estimate the expected uncertainties in our half-life model is to find the fraction of the total lifetime distribution in the extended tail fit by the Secondary Gaussian at long encounter durations.  This corresponds to the fraction of simulations in the lifetime distribution \textit{not} accounted for by our simple half-life formalism.  In order for this method to apply, we assume that each of the $2+2$ and $2+1+1$ outcomes is fit perfectly by the Primary and Secondary curves \emph{in tandem}. With this assumption in mind, we iterate this procedure for all $2+1+1$ and $2+2$ outcomes with the following:\\

For each of the $2+1+1$ and $2+2$ lifetime distributions, let the improper integral from the peak (midpoint) of the Gaussian to $\infty$ be defined as $A_{\rm kP}$ for the Primary Gaussian and $A_{\rm kS}$ for the Secondary Gaussian. After removing the normalization on the y-axis, the integrals thus correspond to the total number of simulations covered by each function on that domain. We let $\epsilon_{\rm k}$ represent the fraction of the total lifetime distribution \textit{not} accounted for by our Primary Gaussian, and hence we use this parameter as an indicator of the uncertainty in our fitting procedure. Then, considering that both Primary and Secondary curves for each outcome have some region $A_{\rm kO}$ that overlaps, to mitigate for double-counting of the same area of the overlap, it follows:

\begin{equation}
\epsilon_k = 1-\frac{A_{kP}}{(A_{kP}+A_{kS})-A_{kO}}.
\end{equation}

\noindent As mentioned before, the $3+1$ outcome did not call for such a calculation, since only a single Gaussian was needed. The calculated values of $\epsilon_{\rm k}$ for the $2+1+1$ and $2+2$ outcomes are included in the right-most column of Table~\ref{table:table}.

In Part I of this series, we used hyperbolic-trigonometric decay models of fit over the cumulative lifetime distributions.  Here, we instead model the lifetime distributions directly.  This is because much of the extended tail seen in the cumulative lifetime distributions can be attributed to the $3+1$ outcome, which is poorly described by our previous fitting function. Consequently, we accept the axiom that the free parameter $\sigma_k$ in Equation~\ref{eqn:Gaussian}, as obtained directly from our fitting functions, is proportional to the half-life.
Hence, fitting to the simulated lifetime distributions directly yields a \textit{proxy} for the half-life $\tau_{k, 1/2}$ of the distribution.  This is a reasonable assumption since the width of the lifetime distribution can be regarded as a proxy for the rate of disruption of four-body systems to a given outcome state.  Hence, the width $\sigma_k$ should be approximately proportional to the corresponding outcome half-life.

Figures \ref{fig:3Models} and \ref{fig:5Models} of the Appendix are similar to Figure \ref{fig:10Models} except that the mass ratio changes, respectively, to $m_{\rm 1}/m_{\rm 2} = 3$ and $m_{\rm 1}/m_{\rm 2} = 5$. An important aspect shared among all three of these graphs is that the red ($2+2$) and blue ($2+1+1$) curves dominate at short disruption times, while the green ($3+1$) dominates at larger values of $\log(\tau_{\rm d}/\tau_{0})$.  We note that the green ($3+1$) histogram corresponds to longer encounter durations, relative to the $2+2$ and $2+1+1$ outcomes, since this outcome requires that one of the four interacting particles ends up on a prolonged excursion far from the system centre of mass, before the single star is ejected \citep{leigh16c}.  This is not the case for the $2+2$ and $2+1+1$ outcomes, which correspond to prompt ejections.

In the subsequent sections, we group together all mass combinations yielding one of our three outcomes when calculating the half-lives.  This is to accommodate our original assumption in \textsection~\ref{method} of indistinguishable particles.  However, we note that this assumption is reasonable since, in each case, a specific mass combination significantly dominates the relative outcome fractions (see \textsection~\ref{masscombs}).  Hence, the corresponding absolute outcome fractions are typically dominated by a specific decay channel, yielding a specific final mass combination.  We show explicitly how the half-lives depend on the distribution of particle masses in the final state in \textsection~\ref{masscombs}.

\subsection{The 2+1+1 outcome} \label{fractions}

Although \texttt{FEWBODY} reports the disruption time of an interacting system of two binaries, it fails to differentiate, in the case of a $2+1+1$ outcome, when the first ejection occurs \citep{fregeau04}. The total system lifetime, including both ejection events, is what is reported by \texttt{FEWBODY}.  In order to calculate the branching ratios or relative fractions for the different decay products, we hypothesize that the time of the first ejection event is the one that is relevant.  This way, all half-lives correspond to the time for the first ejection event to occur for all three decay products, since the 2+2 and 3+1 outcomes have only a single ejection event.  Thus, to fix the lifetime of the $2+1+1$ event by accounting for only the first ejection, we perform the following:\\

Let the total half-life obtained from the $2+1+1$ models of fit in Figures \ref{fig:10Models}, \ref{fig:3Models}, and \ref{fig:5Models} be represented as $\sigma_{\rm C}$, and be related to the half-lives of the first---$\sigma_1$--- and second---$\sigma_2$---ejections by:

\begin{equation}
\label{eqn:sigmasums}
\sigma_{\rm C}=\sigma_1+\sigma_2.
\end{equation}
Then, we make the assumption that:
\begin{equation}
\label{eqn:sigmarelations}
\sigma_1 = \beta \cdot \sigma_2  \mid  \beta \in (0,1),
\end{equation}
where $\beta$ is a constant. From this it follows, through substitution and simplification, that the corrected half-life, denoted as $\sigma'_{\rm C}$ and equivalent to the half-life of the first ejection event $\sigma_1$ is:
\begin{equation}
\label{eqn:kfactor}
\sigma'_{\rm C} = \sigma_1 = \frac{\beta}{1+\beta}\cdot \sigma_{\rm C},
\end{equation}
For consistency, we also set $\sigma'_{\rm A} = \sigma_{\rm A}$ and $\sigma'_{\rm B} = \sigma_{\rm B}$, since the $2+2$ and $3+1$ outcomes have only one ejection event each and do not require correction.  Thus, with $\sigma'_{\rm C}$ representing the true 2+1+1 half-life, the frequency of half-lives that occur per unit time is generically expressed as:
\begin{equation}
\label{eqn:frequency}
\Lambda\equiv\frac{1}{\sigma'},
\end{equation}
from which we can find the relative frequencies of half-lives $\Gamma_k$ as shown in Figures \ref{fig:10Fractions}, \ref{fig:3Fractions}, and \ref{fig:5Fractions} for an arbitrary resultant $k$ out of the total resultant set \{A, B, C\} that represents each of the $2+2, 3+1$, and $2+1+1$ outcomes by:
\begin{equation}
\label{eqn:fractions}
\Gamma_k= \frac{{\sigma'}_k^{-1}}{{\sigma'}_{\rm A}^{-1}+{\sigma'}_{\rm B}^{-1}+{\sigma'}_{\rm C}^{-1}}=\frac{\Lambda_k}{\Lambda_{\rm A}+\Lambda_{\rm B}+\Lambda_{\rm C}}.
\end{equation}

We note that only by calculating and implementing the value of the parameter $\beta$ in Equation~\ref{eqn:kfactor} do we obtain agreement between our half-life formalism and the simulated outcome fractions. 
This is ultimately done to obtain an expectation for the ratio of the two ejection times in the $2+1+1$ outcome needed to obtain this excellent agreement, via the calculated parameter $\beta$.  In \textsection~\ref{confirmation}, we assess this prediction directly using additional numerical scattering simulations that explicitly calculate the time corresponding to the first ejection event in the $2+1+1$ outcome.

We also note that the relative values of the time corresponding to the initial (left-most) bin in Figure~\ref{fig:10Models} for each of the three generic outcomes is consistently proportional to the ratio of their $\sigma'$ values
, which is our proxy for the half-life.  Hence, although it is perhaps sensible to include this initial delay time before the first disruption occurs in our proxy for the half-life, the fortuitous proportionality between this initial time and $\sigma'$ allows us to use only one of these parameters as our proxy for the half-life.

\subsection{Probabilistic distributions for all decay products} \label{probmethod}

Though Figures \ref{fig:10Models}, \ref{fig:3Models} and \ref{fig:5Models} accurately portray the distribution of lifetimes per resultant, the fractions that describe each histogram are relative to that resultant only. Thus, in an attempt to standardize all resultants from each window into fractions relative to one another, we perform this step, which directly yields the probabilities of disruption of specific resultants as a function of time.

If we were to consider a probabilistic point of view using the models of fit, we obtain the \emph{PDFs} for each resultant for each mass ratio $m_{\rm 1}/m_{\rm 2}$. As per the property of \emph{PDFs}, we obtain:
\begin{align}
\label{eqn:probsweep}
P_{k}(t_1\leq\tau_{\rm d}\leq t_2)\nonumber\\
 \equiv &\int_{\log (t_1/\tau_{0})}^{\log(t_2/\tau_{0})} f_{k}(\log(\tau_{\rm d}/\tau_{0}))~d(\log(\tau'_{\rm d}/\tau_{0})).
\end{align}
As $t_{\rm 2} \to t_{\rm 1}$, the \emph{relative} probability of yielding a specific resultant $k$ out of the set of resultants {A, B, C} at a certain disruption time 
 $\tau_{\rm d}$ becomes:

\begin{equation}
\label{eqn:specificprob}
P_{k}(\log(\tau_{\rm d}/\tau_{0})) = \frac{N_{k} \cdot f_{k}(\log(\tau_{\rm d}/\tau_{0}))}{\sum\limits_{{k}={\rm A,B,C}}N_{k} \cdot f_{k}(\log(\tau_{\rm d}/\tau_{0}))}
\end{equation}
where $N_{k}$ represents the total number of simulations, out of 4 $\cdot$ 10$^4$
, that pertain to a specific resultant. 

\subsection{Branching ratios} \label{results}

In this section, we present the simulated fractions for each outcome, for all mass ratios, and compare the results to the half-lives reported in Table~\ref{table:table}.  We emphasize that we are using the $\sigma_{\rm k}$ value measured directly from our fits to the simulated lifetime distribution as a proxy for the true half-life for each outcome.  
We hypothesize that, in analogy with radioactive decay, the ratios of half-lives should be proportional to the relative outcome fractions.  With the half-lives (indirectly) measured, we can test this hypothesis directly using the results of our numerical scattering simulations.  If our half-life formalism is valid, then the ratio of half-lives should be equivalent to the ratio of outcome fractions, or branching ratios. We test this idea directly in this section. 
 
\begin{figure}
\begin{center}                                                                                                                                                           
\includegraphics[width=\columnwidth]{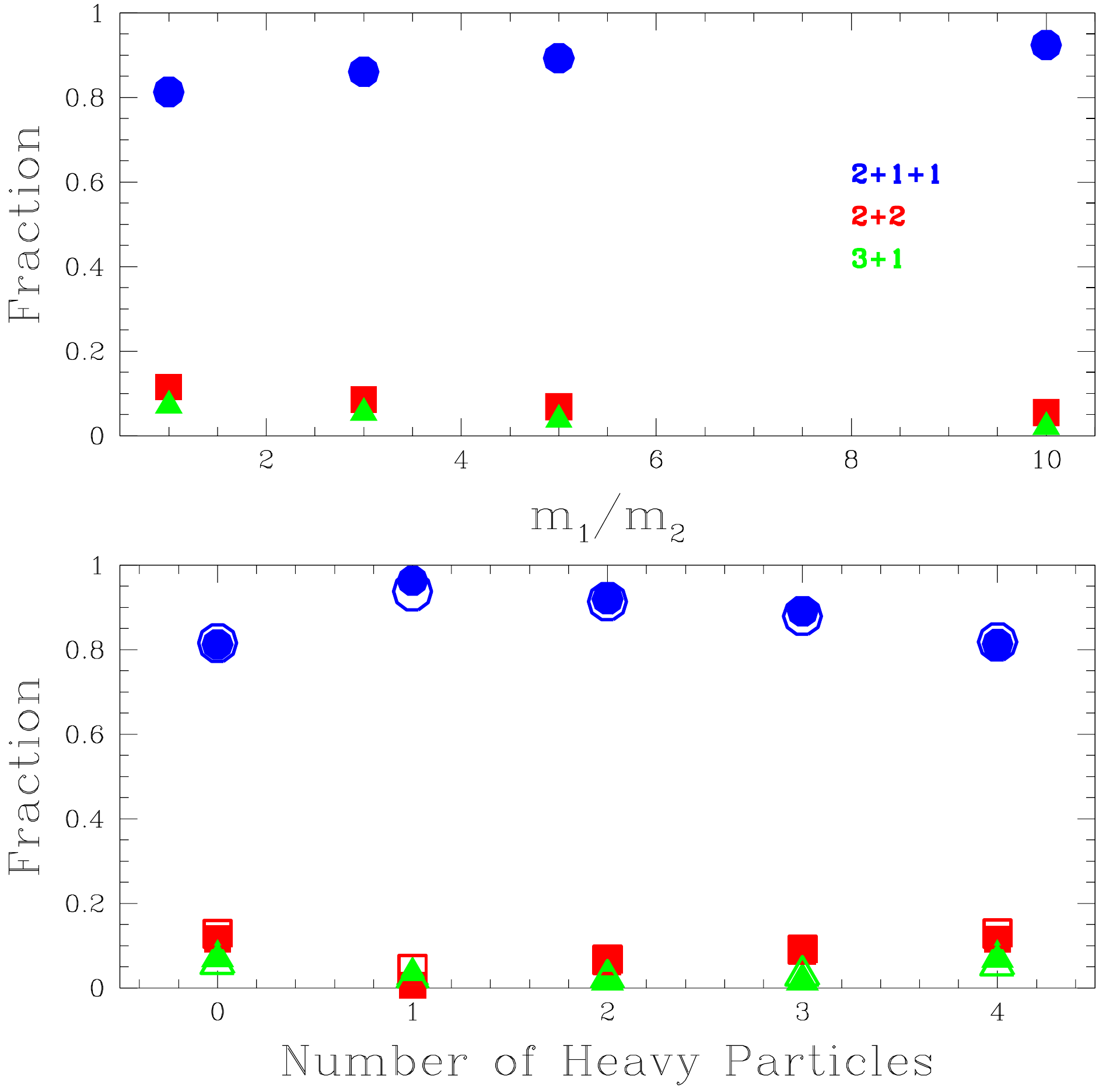}
\end{center}
\caption[The fractions of binary-binary interactions that produce 3+1, 2+2 and 2+1+1 outcomes, as a function of the distributions of particle masses, with focus on $m_1=10\Msol$]{Plot showing the fractions (branching ratios) of simulated binary-binary interactions that produce each of the 3+1, 2+2 and 2+1+1 outcomes, as a function of the number of most-massive (if any) particles.
Here, we fix the particle masses to be either $m_{\rm 1} = 10\Msol$ or $m_{\rm 2} = 1 \Msol$, but vary their relative numbers. We define heavy particles as those above $1 \Msol$, and, in this case, the number of bodies with a mass of $10 \Msol$. The shaded blue circles, red squares, and green triangles show, respectively, the relative fractions of simulations for $2+1+1, 2+2,$ and $3+1$ resultants. The open shapes superimposed on their corresponding shaded shapes represent the relative fractions predicted from the half-lives $\tau_{k, 1/2}$ in \textsection~\ref{virial}. Notice the agreement between the relative fractions of each outcome and those predicted from our half-life formalism, for all combinations of particle masses, but assuming the values for $\beta$ given in Table~\ref{table:table}. The additional simulations performed in Section~\ref{confirmation} will confirm that this close agreement is real.}

\label{fig:10Fractions}
\end{figure}

Table~\ref{table:table} lists the values of the free-parameter constants used to create the models of best fit as seen in Figures~\ref{fig:10Models}, \ref{fig:3Models}, and \ref{fig:5Models}. Some entries are void because the corresponding free parameters do not apply to those resultants. The $\sigma'$ and $\sigma$ values from this table were used to obtain the superimposed curves as seen in Figures \ref{fig:10Fractions}, \ref{fig:3Fractions}, and \ref{fig:5Fractions}. 

We re-iterate that, according to Equation~\ref{eqn:Gaussian}, two fitting parameters are needed to fully characterize the lifetime distributions for each outcome, namely $\sigma_{\rm k}$ and $t_{\rm k,0}$.  However, as already discussed, the latter parameter is proportional to the former.  Thus, we consider only $\sigma_{\rm k}$ in calculating our proxy for the half-life via Equations~\ref{eqn:comb5} and~\ref{eqn:comb6}.

\begin{table*}
\centering
\small
\addtolength{\tabcolsep}{-3pt}
\caption{Values of the free parameters in Equation~\ref{eqn:Gaussian} for all outcomes and mass ratios. Note that the fitting constants bifurcate into the two Gaussian models used for each of the $2+1+1$ and $2+2$ resultants---the Primary that covers most of the simulations, and the Secondary that fits the extended tail. The final column denotes our estimate for the uncertainty as outlined in \textsection~\ref{fittingmethod}.}
\begin{tabular}{ccccccccccccccc}
\toprule
		&		&		&	\vline		&	\multicolumn{5}{c}{\emph{Primary Gaussian}}	&	\vline		&		\multicolumn{3}{c}{\emph{Secondary Gaussian}}	&	\vline	&	\multicolumn{1}{c}{\emph{Uncertainty Proxy}}	\\
\hline
Particle      &  Mass   &  Resultant &	\vline    &	$C$    &    $\log(t_{\rm 0}/\tau_0)$    &    $\sigma/\tau_{0}$	&	$\beta$  & $\sigma'/\tau_0$  &	\vline 	&	$C$ 	& 		$\log(t_{\rm 0}/\tau_0)$	&	$\sigma/\tau_0$	&	\vline	&	$\epsilon$   \\
Masses               &     Combination     &     &           &       &       & 	&	&	&	&	&	&	&	&	\\
\toprule
	&	$N_{\rm 1} =$ 4; $N_{\rm 2} =$ 0  &  $2+2$	&	&	0.80	&	3.05	&	0.32	&	---	&	--- 	&	&	1.88	&	3.05	&	0.75		&	&	0.28	\\
	&	&	$2+1+1$	&	&	0.50	&	2.88	&	0.20	&	0.35	&	0.05			&	&	0.88	&	2.88	&	0.35	&	&	0.21	\\
	&	&	$3+1$	&	&	1.88	&	4.50	&	0.75	&	---	&	---	&	&	---	&	---	&	---	&	&	---	\\
	&    $N_{\rm 1} =$ 3; $N_{\rm 2} =$ 1  &  $2+2$   &	&  	0.69     &	3.10	&	0.28		& ---	& --- &	&	1.50	&	3.10  	&	0.60		&	&	0.27 \\
	&	&	$2+1+1$	&	&	0.50	&	3.00	&	0.20	&	0.18	&	0.03		&	&	0.88	&	3.00	&	0.35		&	&	0.21	\\
	&	&	$3+1$	&	&	2.01	&	4.50	&	0.80	&	---	&	---	&	&	---	&	---	&	---	&	&	---	\\
$m_{\rm 1} =3 \Msol$	&    $N_{\rm 1} =$ 2; $N_{\rm 2} =$ 2  & $2+2$    &   &	0.60    &  3.10	&	0.24	&	---	&	---	&	&	1.50	&	3.10	&	0.60	&	&	0.29	\\
	&	&	$2+1+1$	&	&	0.63	&	3.05	&	0.25	&	0.12	&	0.03		&	&	1.13	&	3.05	&	0.45		&	&	0.22\\
$m_{\rm 2} =1\Msol$	&	&	$3+1$	&	&	1.75	&	4.30	&	0.70	&	---	&	---	&	&	---	&	---	&	---		&	&	---	\\
	&    $N_{\rm 1} =$ 1; $N_{\rm 2} =$ 3  &  $2+2$    &	&    0.88   &	3.35	&	0.35	&	---	&	---    &	&	1.50	&	3.35	&	0.60		&	&	0.20 \\
	&	&	$2+1+1$	&	&	0.45	&	3.10	&	0.18	&	0.15	&	0.02		&	&	0.75	&	3.10	&	0.30		&	&	0.19	\\
	&	&	$3+1$	&	&	1.75	&	4.60	&	0.70	&	---	&	---	&	&	---	&	---	&	---	&	&	---	\\
	&    $N_{\rm 1} =$ 0; $N_{\rm 2} =$ 4  &  $2+2$    & 	&	0.83      &	3.30	&	0.33	&	---	&	---   &	&	1.88	&	3.30	&	0.75	&	&	0.23	\\
	&	&	$2+1+1$	&	&	0.50	&	3.10	&	0.20	&	0.35	&	0.05		&	&	1.00	&	3.10	&	0.40		&	&	0.24	\\
	&	&	$3+1$	&	&	1.88	&	4.80	&	0.75	&	---	&	---	&	&	---	&	---	&	---	&	&	---	\\
\hline
	&    $N_{\rm 1} =$ 4; $N_{\rm 2} =$ 0  &  $2+2$   	&	& 	0.85      &		2.95	&	0.34	&	---	&	---   &	&	2.01	&	2.95	&	0.80		&	&	0.34 \\
	&	&	$2+1+1$	&	&	0.50	&	2.75	&	0.20	&	0.35	&	0.05	&	&	0.95	&	2.75	&	0.38		&	&	0.23	\\
	&	&	$3+1$	&	&	1.75	&	4.40	&	0.70	&	---	&	---	&	&	---	&	---	&	---	&	&	---	\\
	&    $N_{\rm 1} =$ 3; $N_{\rm 2} =$ 1  & $2+2$      &	    &	   0.68       &		3.00	&	0.27	&	---	&	---    &	&	1.38	&	3.00	&	0.55	&	&	0.25	\\
	&	&	$2+1+1$	&	&	0.48	&	2.92	&	0.19	&	0.20	&	0.03		&	&	0.88	&	2.92	&	0.35		&	&	0.24	\\
	&	&	$3+1$	&	&	2.13	&	4.40	&	0.85	&	---	&	---	&	&	---	&	---	&	---	&	&	---	\\
$m_{\rm 1} =5 \Msol$	&    $N_{\rm 1} =$ 2; $N_{\rm 2} =$ 2  & $2+2$   &	&	0.63       &	 3.15	&	0.25	&	---	&	---  &		&	1.30	&	3.15	&	0.52	&	&	0.25	\\
	&	&	$2+1+1$	&	&	0.75	&	3.00	&	0.30	&	0.10	&	0.03		&	&	1.38	&	3.00	&	0.55		&	&	0.22	\\
$m_{\rm 2} = 1 \Msol$	&	&	$3+1$	&	&	2.01	&	4.20	&	0.80	&	---	&	---	&	&	---	&	---	&	---	&	&	---	\\
	&    $N_{\rm 1} =$ 1; $N_{\rm 2} =$ 3  &   $2+2$   &   	&	1.25       &		3.40	&	0.50	&	---	&	---   &	&	3.26	&	3.40	&	1.30		&	&	0.30	\\
	&	&	$2+1+1$	&	&	0.48	&	3.10	&	0.19	&	0.22	&	0.03		&	&	0.88	&	3.10	&	0.35	&	&	0.22		\\
	&	&	$3+1$	&	&	2.01	&	4.60	&	0.80	&	---	&	---	&	&	---	&	---	&	---	&	&	---	\\
	&    $N_{\rm 1} =$ 0; $N_{\rm 2} =$ 4  &  $2+2$   &    	&	0.83      &	3.30	&	0.33	&	---	&	---   &	&	1.88	&	3.30	&	0.75	&	&	0.23	\\
	&	&	$2+1+1$	&	&	0.50	&	3.10	&	0.20	&	0.35	&	0.05		&	&	1.00	&	3.10	&	0.40		&	&	0.24 \\
	&	&	$3+1$	&	&	1.88	&	4.80	&	0.75	&	---	&	---	&	&	---	&	---	&	---	&	&	---	\\
\hline
	&    $N_{\rm 1} =$ 4; $N_{\rm 2} =$ 0    &   $2+2$  &   	&	0.83      &  2.80	&	0.33	&	---	&	---  &		&	1.75	&	2.80 	&	0.70		&	&	0.26 \\
	&	&	$2+1+1$	&	&	0.50	&	2.60	&	0.20	&	0.35	&	0.05		&	&	1.00	&	2.60	&	0.40	&	&	0.24	\\
	&	&	$3+1$	&	&	2.01	&	4.30	&	0.80	&	---	&	---	&	&	---	&	---	&	---		&	&	---	\\
	&    $N_{\rm 1} =$ 3; $N_{\rm 2} =$ 1  &   $2+2$    &  	&	0.65       & 	2.80	&	0.26	&	---	&	---    &	&	1.38	&	2.80	&	0.55	&	&	0.26	\\
	&	&	$2+1+1$	&	&	0.42	&	2.78	&	0.17	&	0.18	&	0.03		&	&	0.88	&	2.78	&	0.35	&	&	0.25	\\
	&	&	$3+1$	&	&	2.01	&	4.50	&	0.80	&	---	&	---	&	&	---	&	---	&	---	&	&	---	\\
$m_{\rm 1} = 10 \Msol$	&    $N_{\rm 1} =$ 2; $N_{\rm 2} =$ 2  &  $2+2$   	 &      	&	0.68       &	3.25	&	0.27	&	---	&	---  &		&	1.00	&	3.25	&	0.40		&	&	0.26 \\
	&	&	$2+1+1$	&	&	0.75	&	2.65	&	0.30	&	0.07	&	0.02		&	&	1.88	&	3.10	&	0.75	&	&	0.39		\\
$m_{\rm 2} = 1 \Msol$	&	&	$3+1$	&	&	2.26	&	4.50	&	0.90	&	---	&	---	&	&	---	&	---	&	---	&	&	---	\\
	&    $N_{\rm 1} =$ 1; $N_{\rm 2} =$ 3 &   $2+2$     &  	&	1.00       &	 3.20 	&	0.40	&	---	&	---   &	&	3.01	&	3.20	&	1.20	&	&	0.33	\\
	&	&	$2+1+1$	&	&	0.53	&	3.10	&	0.21	&	0.10	&	0.02	&	&	1.00	&	3.10	&	0.40		&	&	0.23	\\
	&	&	$3+1$	&	&	2.51	&	5.00	&	1.00	&	---	&	---	&	&	---	&	---	&	---	&	&	---	\\
	&    $N_{\rm 1} =$ 0; $N_{\rm 2} =$ 4  &  $2+2$    &   	&	0.83      &	3.30	&	0.33	&	---	&	---   &	&	1.88	&	3.30	&	0.75	&	&	0.23	\\
	&	&	$2+1+1$	&	&	0.50	&	3.10	&	0.20	&	0.35	&	0.05		&	&	1.00	&	3.10	&	0.40		&	&	0.24	\\
	&	&	$3+1$	&	&	1.88	&	4.80	&	0.75	&	---	&	---	&	&	---	&	---	&	---		&	&	---	\\
\bottomrule
\label{table:table}
\end{tabular}
\end{table*}

The fractions of binary-binary interactions that produce each of the $3+1$, $2+2$ and $2+1+1$ outcomes are shown in Figure~\ref{fig:10Fractions}, as a function of the distributions of particle masses.  For each number of heavy particles, we select different values of $\beta$ as provided in Table~\ref{table:table}.  In addition, we fix the particle masses to be either $m_{\rm 1} = 10 \Msol$ or $m_{\rm 2} = 1 \Msol$, but vary their relative numbers.  The outcome fractions are shown as a function of the number of heavy (i.e., $m_{\rm 1} = 10 \Msol$) particles. The blue circles, red squares, and green triangles represent, respectively, the $2+1+1, 2+2$ and $3+1$ outcomes. Figures \ref{fig:3Fractions} and \ref{fig:5Fractions} in the Appendix are similar to Figure~\ref{fig:10Fractions} but show the results for mass ratios of $m_{\rm 1}/m_{\rm 2} = 3$ and $m_{\rm 1}/m_{\rm 2} = 5$, respectively. Notice the close agreement between the superimposed open symbols that were obtained from the half-life method described in \textsection~\ref{virial} and the solid symbols obtained directly from the experimental simulations.  We emphasize that, provided the simulated ratio of disruption times between the first and second ejection events in the $2+1+1$ outcome are as calculated in Figure~\ref{fig:10Fractions}, then this close agreement will reflect the actual simulated results.  We will test this directly in the next section.

\begin{figure}
\begin{center}
\includegraphics[width=\columnwidth]{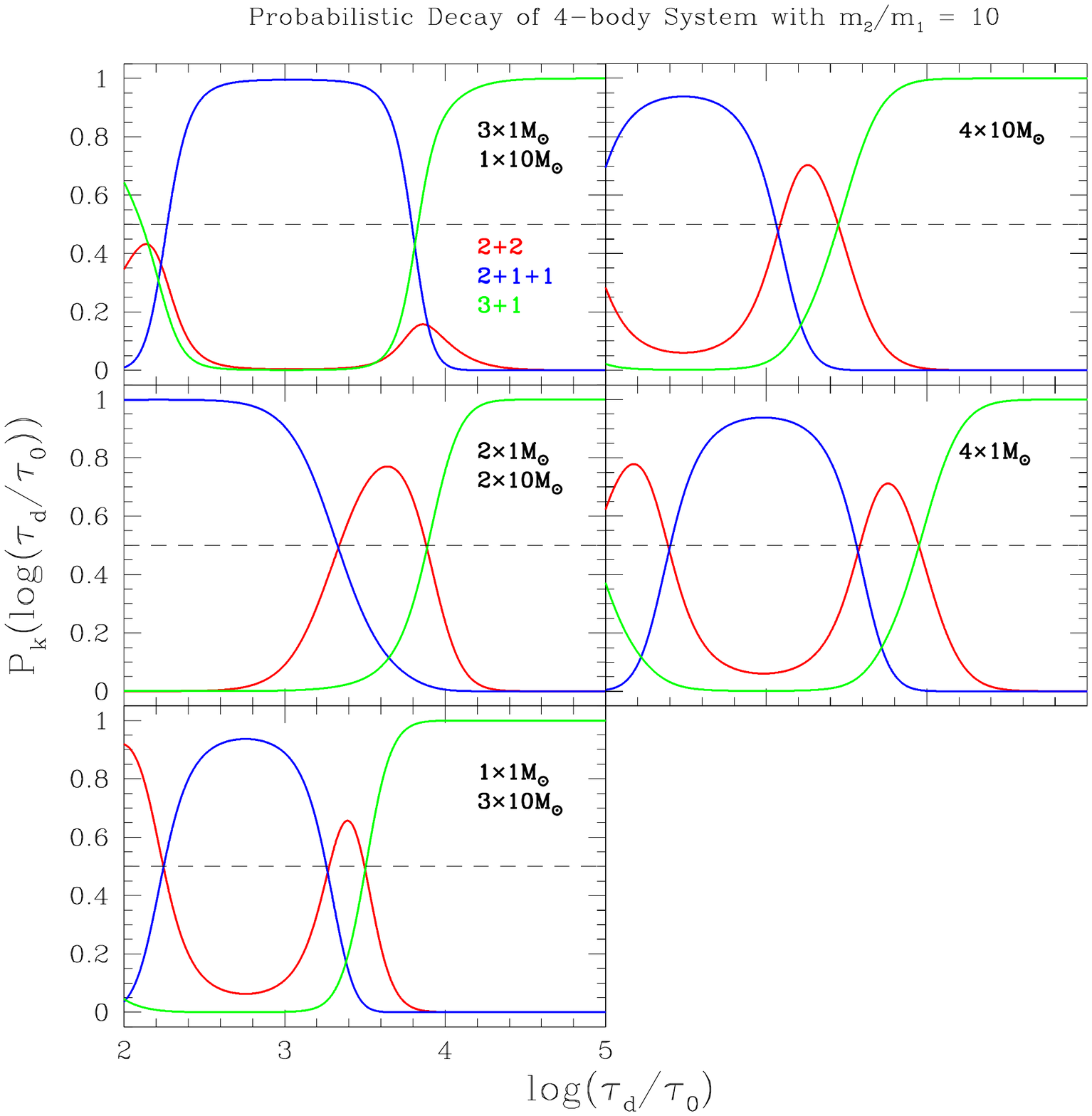}
\end{center}
\caption{The probability of decay into a resultant as time progresses for $m_1=10\Msol$. The red, blue, and green curves correspond to $2+2,~2+1+1,~3+1$ outcomes, respectively. The horizontal dashed line is an auxiliary feature used to indicate half-lives on this probability graph. 
Notice that though the $x$-axis represents disruption time, it is also the progression of time. At every $x$-coordinate, the sum of probabilities from all resultants is, of course, unity.}
\label{fig:10Probabilistic}
\end{figure}

Using the method described in \textsection~\ref{probmethod}, we obtain Figures
 \ref{fig:10Probabilistic}, \ref{fig:3Probabilistic} and \ref{fig:5Probabilistic} for the mass ratios $m_{\rm 1}/m_{\rm 2} = 10, 3$ and $5$, respectively. These figures show the instantaneous relative disruption rates for all three encounter outcomes, as a function of the total encounter duration or time.  There are several interesting patterns that can be observed from these plots.  
For example, the general alteration pattern between dominations is from $3+1$ to $2+2$ to $2+1+1$ to $2+2$ and finally to $3+1$, the latter of which is completely expected due to the presence of a long-period excursion immediately prior to the only ejection event \citep{leigh16b}.
The $2\times1\Msol$ plots in all the aforementioned probabilistic figures seem most distinct from the other mass combinations. This may be due to these plots having the biggest number of possible combinations of how the two $1\Msol$ and the two $3~\Msol, 5~\Msol$, or $10~\Msol$ particles can form final states in binary-binary interactions.

In addition, though the equations $f_{k}$ explained in \textsection~\ref{fittingmethod} are defined over the $\log(\tau_{\rm d}/\tau_{0}) \in \mathbbm{R}$ interval of the abscissas, the interval over which the interaction of each resultant's probability is present is on the restricted domain from $\log(\tau_{\rm d}/\tau_{0}) = 2$ to $\log(\tau_{\rm d}/\tau_{0}) = 5$, after which the $3+1$ outcome reigns.

\subsection{Ejection time distributions for $2+1+1$}
\label{confirmation}

The $\beta$ values in Table~\ref{table:table} should be equal to the ratio of half-lives between the first and second ejection events of a 2+1+1 outcome.  That is, these are the ratios needed in order for our half-life formalism to provide the best possible agreement with the simulated outcome fractions.  Next, we test this hypothesis using additional simulations that take into account the time of the first ejection in the $2+1+1$ outcome.  If this test confirms our hypothesis, this will be a strong indicator of the accuracy of our model.  

\begin{figure}
\begin{center}
\includegraphics[width=\columnwidth]{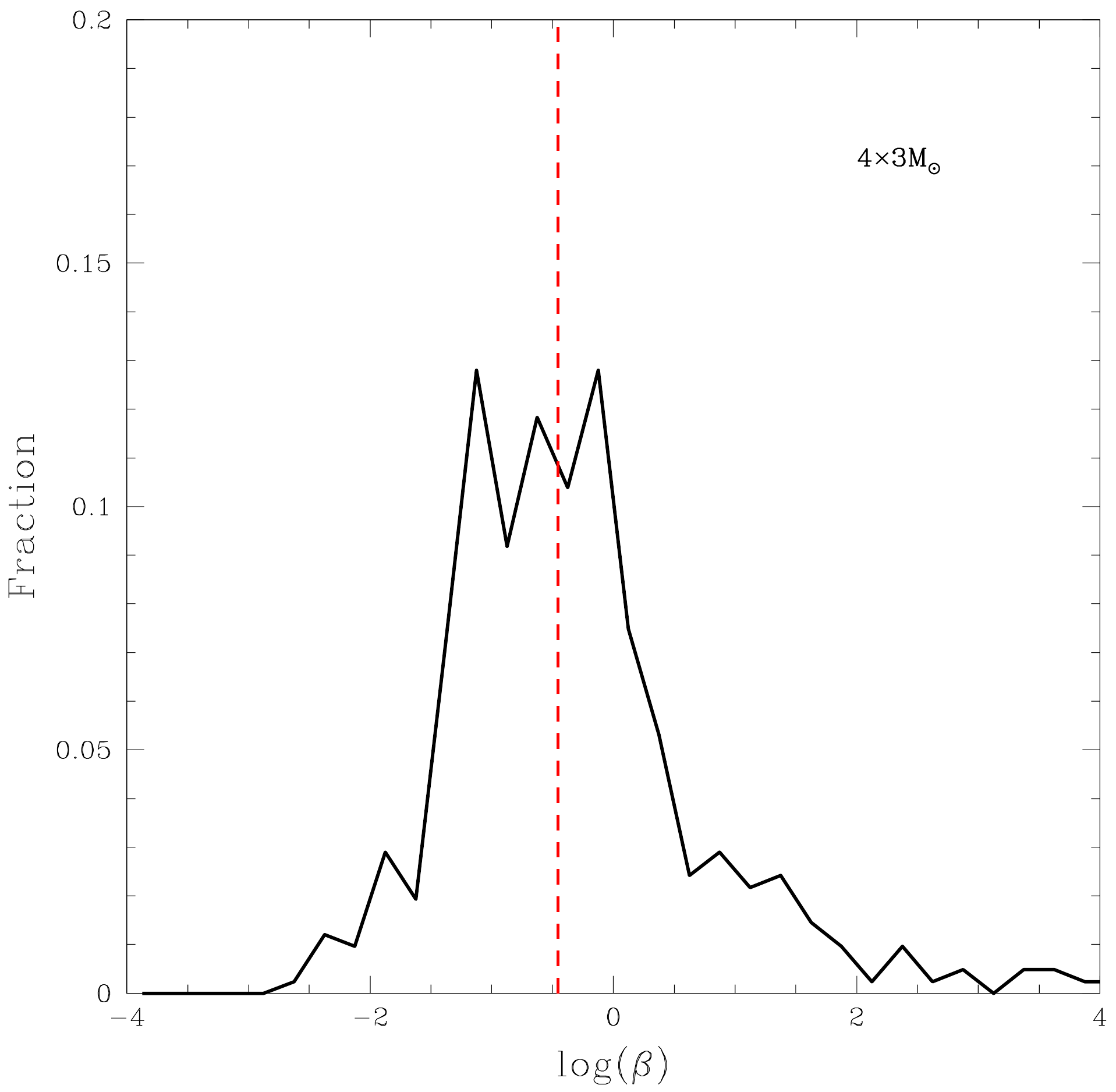}
\end{center}
\caption{The distribution of $\beta$' values for the 414 scattering simulations performed with the code from \citet{ryu16} and \citet{ryu17a}, as described in the text.  The solid black histogram shows the results for the case $N_{\rm 1} =$ 4; $N_{\rm 2} =$ 0 with $m_{\rm 1}/m_{\rm 2} = 3$.  For comparison, we show by the vertical, dashed red line the corresponding predicted value for $\beta$ from Table~\ref{table:table}, i.e. $\log{\beta}=-0.456$}. Note that the previous calculated value of $\beta$ aligns neatly at the median and roughly at the mean of the $\beta'$ distribution. 
\label{fig:fig7}
\end{figure}

Due to limitations in computational resources, our allotted time to perform the additional simulations was limited.  We obtained an additional 414 simulations of binary-binary scatterings using the code presented in \citet{ryu16} and \citet{ryu17a,ryu17b,ryu17c}.  As we will show, this turns out to be a sufficiently large number of simulations to accurately test our hypothesis.  This code calculates the times at which each single star is ejected during a 2+1+1 outcome, something \texttt{FEWBODY} is unable to perform.  This is done by evaluating the time-step at which the relative energy between a single star and the remaining substellar system (binary or triple) becomes positive. For these simulations, we adopt the same initial conditions as in \textsection~\ref{exp}, and focus on the case with identical particles each having a mass of $3\Msol$.  In order to perform the necessary comparison, we must subtract from each escape time the initial fall-in time of the binaries, such that the point of direct impact is taken to occur at t $=$ 0.  This initial drop-in time is of order $\sim$ 250 years in our simulations.\footnote{There is some minor variation in the initial drop-in time between our simulations.  As such, we have verified that our results are insensitive to the exact choice for this time, within realistic bounds.}  We then calculate the parameter $\beta$' as:
\begin{equation}
\label{eqn:beta2}
\beta' = \frac{t_{\rm 1}}{t_{\rm 2} - t_{\rm 1}},
\end{equation}
where $t_{\rm 1}$ and $t_{\rm 2}$ are the times corresponding to the first and second escape events, respectively.  The distribution of $\beta$' values for our suite of scattering simulations is shown in Figure~\ref{fig:fig7} by the solid black frequency plot.  For comparison, we show by the vertical dotted red line the predicted value for $\beta$ from Table~\ref{table:table} for the case $N_{\rm 1} =$ 4; $N_{\rm 2} =$ 0 with $m_{\rm 1}/m_{\rm 2} = 3$, i.e. $\log{\beta}=-0.456$.
  As is clear, the agreement between our previously calculated value for $\beta$ is strong with the distribution of values for $\beta$' we get from the new simulations.  Our value for $\beta$ falls very close to the median of the distribution of $\beta$' values we obtain from these additional simulations. It is important to keep in mind that each of the follow-up 414 simulations yielded independent values of $\beta$'. Thus, a distribution plot as presented in Figure~\ref{fig:fig7} is most appropriate to assess our single-value predictions.

\subsection{Half-lives for different mass combinations of the decay products} \label{masscombs}

In this section, we relax our assumption of indistinguishable particles, to better understand how the half-life might depend on the distributions of particle masses.  We consider one such case, selected to best represent all our simulated data, that was partitioned into different combinations of particle masses in the decay products.  We defer to a future paper a more thorough exploration of the various ways that the measured system half-life, or some proxy for it, might depend on the distribution of particle masses.

For the case $2\times1\Msol+2\times5\Msol$, the possible final mass combinations are:

{\renewcommand{\labelitemi}{$\bullet$}
\begin{itemize}
\item $2+2$ outcome:
	\begin{itemize}
	\item HH+LL
	\item HL+HL
	\end{itemize}
\item $2+1+1$ outcome:
	\begin{itemize}
	\item HH+L+L
	\item HL+H+L
	\item LL+H+H
	\end{itemize}
\item $3+1$ outcome:
	\begin{itemize}
	\item HHL+L
	\item HLL+H
	\end{itemize}
\end{itemize}}

where \lq\lq H" represents a heavy particle (in this case $m_{\rm H} = 5\Msol$), and \lq\lq L" a light particle ($m_{\rm L} = 1\Msol$).\footnote{Note that we do not further sub-divide the $3+1$ outcomes according to the exact hierarchical configuration of any triples.}

\begin{figure}
\begin{center}
\includegraphics[width=\columnwidth]{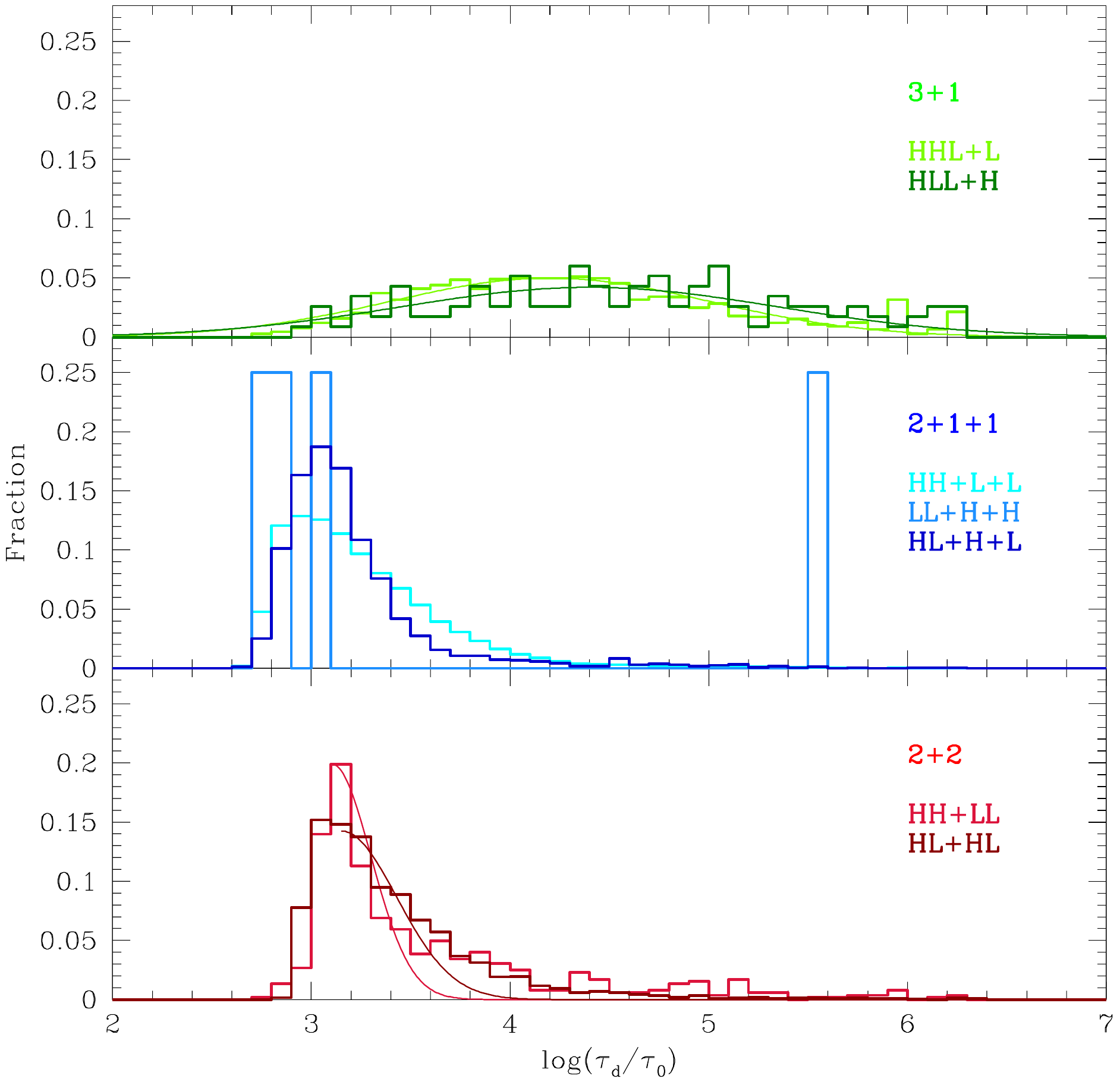}
\end{center}
\caption[Partitioned mass combinations for the decay products]{The different possibilities a $2\times1\Msol + 2\times5\Msol$ system can break up into for each of its three resultants. The possibilities are plotted as lifetime distributions on a normalized y-axis, similar to Figure~\ref{fig:10Models}. Notice how the LL+H+H outcome does not follow a Gaussian-like distribution.  This is because there are so few instances of this outcome occurring (only 4 occurrences out of 40000). For all other cases, we fit a Gaussian curve; the right-halves for the red ($2+2$) resultants, and the full curves for the green ($3+1$) resultants. The blue curves are not fit, because the lifetime distributions correspond to the times for the second ejection events to occur.  As described in \textsection~\ref{confirmation}, we now know that we really want to fit the lifetime distributions for the first ejection events.  Hence, the half-lives obtained from fitting the plotted 2+1+1 lifetime distributions will not yield branching ratios that agree with the simulations. The values of the free parameters obtained from our fits, as given in Equation~\ref{eqn:Gaussian}, are shown in Table~\ref{table:table2}.}
\label{fig:fig8}
\end{figure}

By partitioning the resultants for these mass combinations into the aforementioned categories, we create Figure~\ref{fig:fig8} with superimposed Gaussian fits similar to the procedure in \textsection~\ref{fittingmethod}.  The values of the free parameters obtained from our Gaussian fits, as given in Equation~\ref{eqn:Gaussian}, are shown in in Table~\ref{table:table2}.

\begin{table*}
\centering
\caption{The values of the free parameters used to create the models of fit in Figure~\ref{fig:fig8}. The values for $2+1+1$ are omitted because of the nature of the double ejection which requires inclusion of the $\beta$ parameter, for which we do not have sufficient additional simulations to calculate for all mass combinations considered in this paper. The last column represents the ratio of $\sigma$ values for each outcome, namely the top-most mass type to the bottom-most mass type, as seen in the table.}
\begin{tabular}{ccccccc}

\toprule
Mass Combination	&	Outcome	&	Mass Type	&	$C$	&	$\log(t_0/\tau_0)$	&	$\sigma/\tau_0$	&	Ratio		\\
\toprule
	&	2+2	&		HH+LL	&	0.50	&	3.10	&	0.20		&	0.71 \\
	$2\times1\Msol + 2\times5\Msol$	&		&		HL+HL	&	0.70	&	3.15	&	0.28		&		\\
\cline{2-7}
	&	3+1	&		HHL+L	&	2.01	&	4.20	&	0.80	&	0.84	\\
	&		&		HLL+H	&	2.38	&	4.40	&	0.95	&			\\
\bottomrule
\label{table:table2}
\end{tabular}
\end{table*}

\begin{figure}
\begin{center}
\includegraphics[width=\columnwidth]{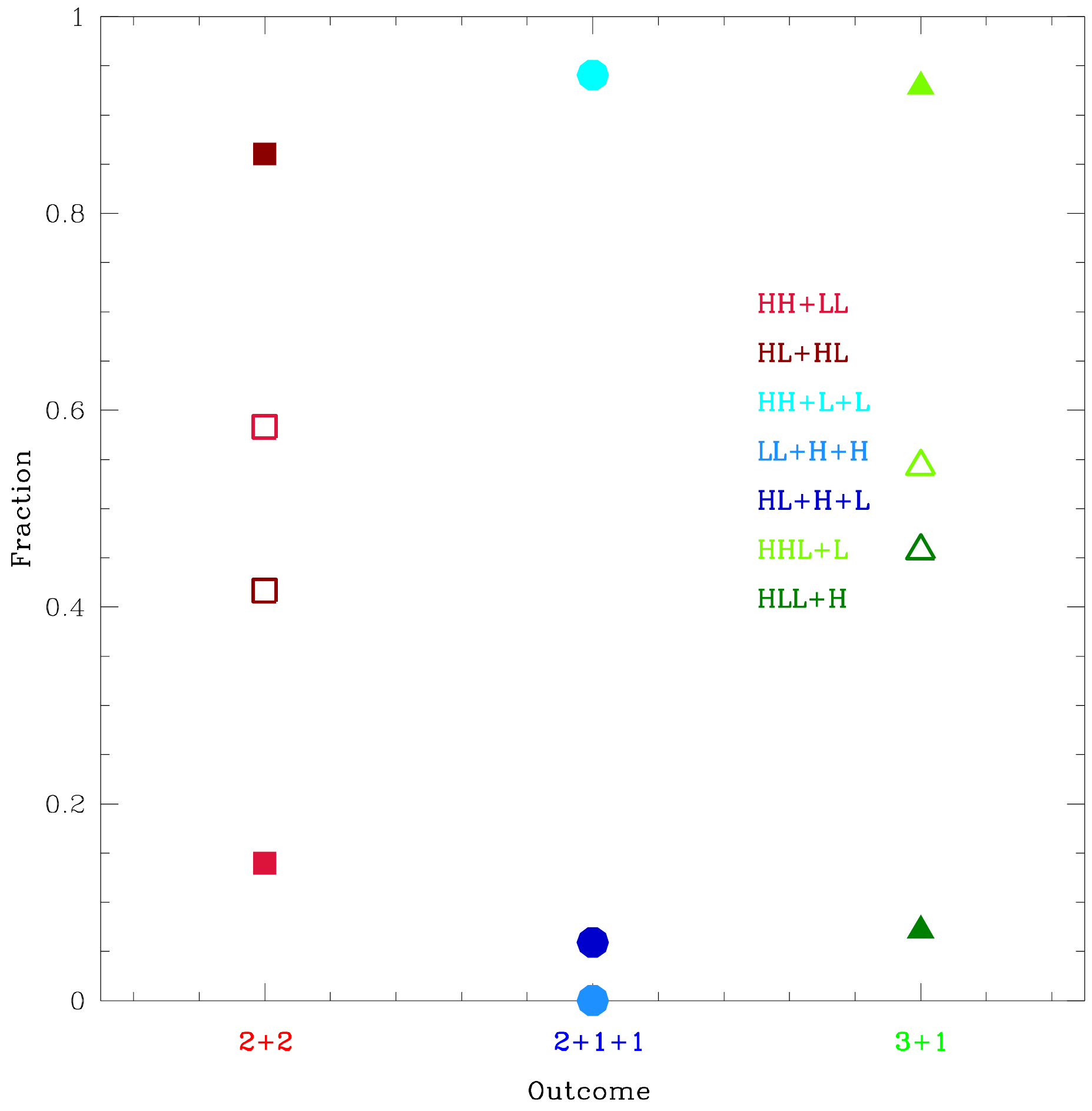}
\end{center}
\caption{Outcome fractions partitioned among the different mass combinations, for the $2\times1\Msol + 2\times5\Msol$ case. Notice how the LL+H+H outcome is exceedingly rare for the $2+1+1$ resultant, with the most common outcome being HH+L+L. In addition, the HL+HL and HHL+L outcomes are dominant for the $2+2$ and $3+1$ resultants, respectively. Just as in Figure~\ref{fig:10Fractions} and its analogs, the filled shapes represent the relative outcome fractions found directly from the simulations, while the open shapes represent the half-life-derived fractions.}
\label{fig:fig9}
\end{figure}

Though Figure \ref{fig:fig8} shows the fraction of simulations that disrupt as a function of time, these fractions are calculated relative to each individual outcome. Recreating Figure~\ref{fig:10Fractions} but using these calculated outcome fractions instead, we obtain Figure~\ref{fig:fig9}.

In Figure \ref{fig:fig9}, we see poor agreement between the calculated outcome fractions and those predicted from our half-life formalism, for which we speculate upon a few possible explanations:
\begin{enumerate}
\item The HH+LL outcome for the $2+2$ case and the HLL+H outcome for the $3+1$ case yield only 523 and 116 simulated occurrences, respectively, out of 40,000 total scattering experiments. Consequently, we do not achieve sufficient statistical significance for these outcomes to safely conclude that the lifetime distributions have converged, and that our Gaussian fits are reliable.  This inhibits us from drawing any firm conclusions from this experiment, since more data is needed to create models of fit.  Moreover, it is not clear to what extent these results depends on our chosen initial conditions (e.g., impact parameter, relative velocity at infinity, etc.), and more work is needed to better understand these dependences.
\item Our simple model for calculating the outcome fractions from the fitted half-lives does not adequately incorporate the underlying physics responsible for generating the simulated lifetime distributions for some mass combinations.  That is, not all mass combinations necessarily correspond to long-lived resonant states, such that the assumption of ergodicity breaks down more severely for these mass combinations.  This is a key requirement in order for our half-life formalism to yield reliable outcome fractions.
\end{enumerate}
In future studies, we intend to adjust our model to account for particle mass, in an attempt to find an agreement for mass-dependent branching ratios.

\section{Discussion} \label{discussion}

In this section, we discuss the underlying assumptions inherent to our method and some of their possible limitations, as well as the astrophysical significance of our results.

\subsection{Assumptions} \label{assumptions}

\begin{figure}
\begin{center}
\includegraphics[width=\columnwidth]{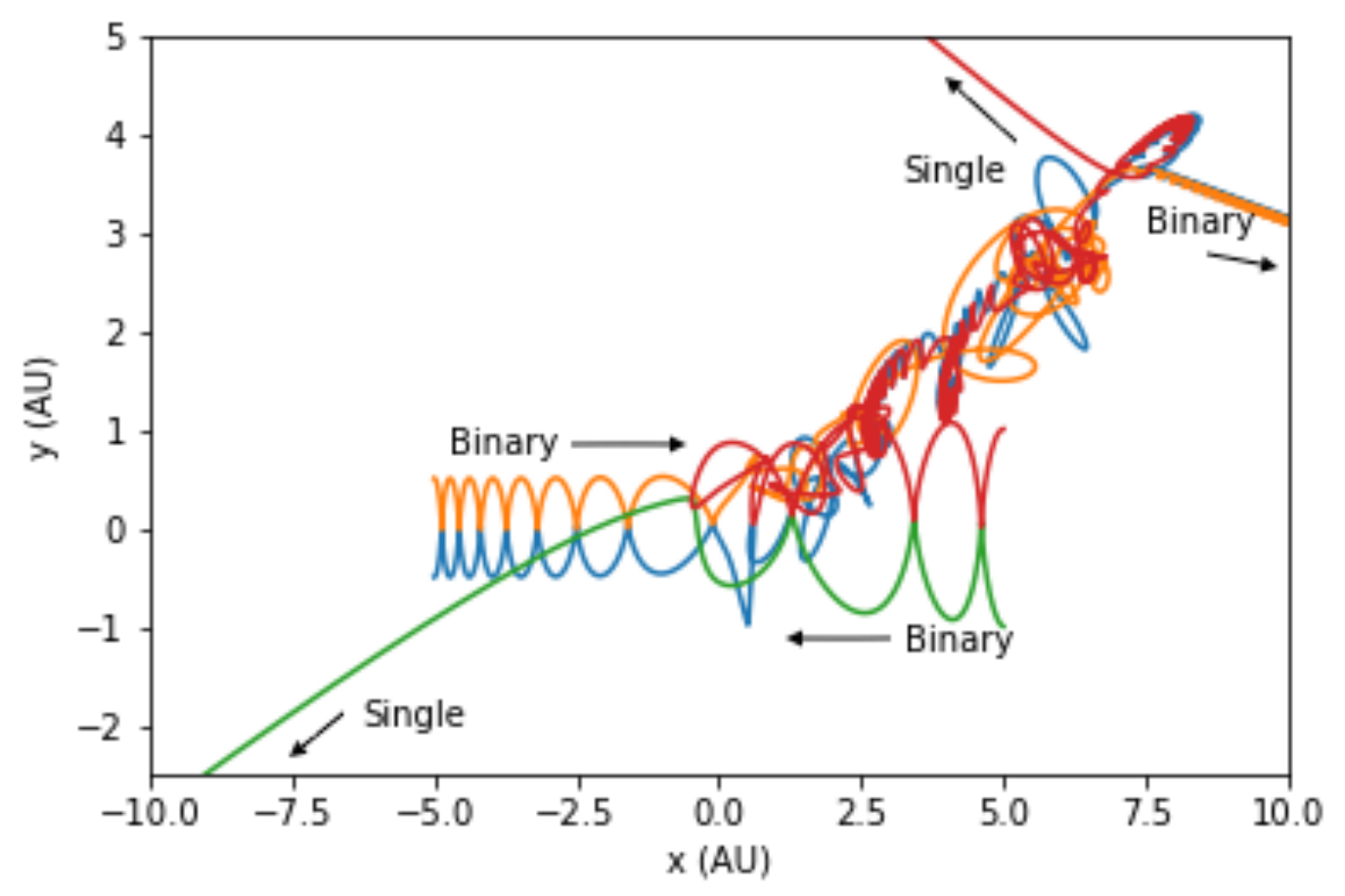}
\end{center}
\caption{The evolution of a binary-binary encounter in position space, projected on to a 2-D plane. All four bodies have equal mass. The distance scale on both axes is in astronomical units.  The two initial binaries come in along the x-axis in opposite directions, and impact at the origin.  The green star is ejected first, almost immediately after direct impact.  A prolonged excursion between the orange, red and blue stars follows, until eventually the red star is ejected, leaving the orange and blue stars in a very compact binary.}
\label{fig:fig10}
\end{figure}

In order to obtain good agreement between the simulated branching ratios (or outcome fractions) and those calculated from our half-life formalism shown in Figures~\ref{fig:10Models}, ~\ref{fig:3Models} and~\ref{fig:5Models}, the derived values for the parameter $\beta$ provided in Table~\ref{table:table} must agree with the corresponding $\beta$' values.  That is, our calculations for the required ratio between successive escape times for both single stars ejected during the 2+1+1 outcome must yield a \lq\lq corrected" half-life for this outcome that is much shorter than the half-lives for the other two outcomes.  This is needed to ensure that the ratio of half-lives for each outcome found from fitting to the simulated lifetime distributions is roughly equal to the ratio of outcome fractions found directly from the simulation.  In this paper, we have demonstrated using a limited number of additional scattering simulations that our calculation for the value of $\beta$ is consistent with the median, and roughly the mean, of the simulated values for $\beta$', but only for one set of initial conditions.  We intend to perform this same comparison for the rest of our simulations in a future paper.  Regardless, we have shown here that a common scenario involves the rapid ejection of the first single star, followed by a prolonged chaotic interaction between the remaining three bodies, before the second star finally escapes.  This generic picture is illustrated in Figure~\ref{fig:fig10} (see also Figure A1 in \citet{leigh12}).

In order to obtain some estimate for the half-life for each encounter outcome, we chose to fit our simulated lifetime distributions with Gaussian distributions, and used the width of each Gaussian as a proxy for the system half-life.  We assume that this provides a good estimate of the time within which a fixed fraction of simulations will have disrupted, for each outcome.  Conversely, in Part I, we worked with the cumulative lifetime distributions and fit for the half-lives directly.  This change proved necessary in the present study since the cumulative lifetime distribution for the 3+1 outcome is poorly described by an exponential curve, and is responsible for much of the extended tails reported in Part I.  This is apparent from a brief glimpse at the lifetime distributions for the 3+1 outcome shown in Figure~\ref{fig:10Models}, which span almost five orders of magnitude.  Although the Gaussian method seems to work well given the results presented in this paper, more work needs to be done to verify the validity of this assumption over a larger range of parameter space.

Our results suggest that our basic method works the best assuming all identical particles.  We speculate that this is due to the assumption of ergodicity being the most accurately upheld in the limit of all equal-mass particles.  However, this might also be related to the fact that our chosen initial conditions are such that smaller mass ratios correspond to larger differences in the total encounter energy between different sets of simulations.  This is because, although we fix the numbers of particles at a given mass in a given set of simulations, which particle ends up with a given mass is chosen at random, and we fix the binary orbital parameters to be constants and the initial relative velocity at infinity to be a constant fraction of the critical velocity.  Thus, the encounter energies are only ever strictly the same for all simulations within a given set (i.e., mass combination) if all particles are identical.  If correct, this suggests that, in future work, the total encounter energy should be held strictly constant when fitting to the simulated lifetime distributions to obtain a more accurate estimate for the half-life.

With that said, more work is needed to better understand the sensitivity of our results to our very limited set of chosen initial conditions.  In particular, we assume a single impact parameter of zero and a single relative velocity at infinity, restricting ourselves to a narrow range of total encounter energies and angular momenta.  Hence, as discussed further in the subsequent section, varying these parameters in future studies should help us to better understand the dependence of the basic formalism and results presented here on our choice of initial total encounter energies and angular momenta.

Finally, we emphasize that we use the Primary Gaussian fits shown in Figures~\ref{fig:10Models}, ~\ref{fig:3Models} and~\ref{fig:5Models} to calculate our estimates for the outcome half-lives.  While the Primary Gaussians cover on average $\approx70\%$ of the total number of simulations, the Secondary Gaussians are needed to fit the remaining distributions at long disruption times.  Consequently, this suggests that our estimates for the branching ratios calculated using our half-life formalism should only be accurate most of the time ($70\%$).  This should be kept in mind in future studies, and other fitting methods should be further experimented with.  With that said, as shown here, the Secondary Gaussians offer one convenient method of estimating uncertainties for the derived half-lives.

\subsection{Initial conditions} \label{ICs}

In this section, we discuss the sensitivity of our results to our chosen initial conditions for our numerical scattering simulations.  We have assumed all circular binaries initially, with a focus on equal initial orbital separations, zero impact parameter and fixed relative velocity at infinity.  We have chosen the relative velocity at infinity using previous works as a guide to the initial encounter energies for which we expect the assumption of ergodicity to be up held \citep{leigh16b,leigh16c}.  Naively, we do not expect our results to follow a half-life formalism in the limit where ergodicity breaks down, since here the particles do not thoroughly mix in phase space, and the interactions tend to be prompt.  With that said, this should be tested more thoroughly in future studies.

We have also not considered the dependence of our results on the total encounter angular momentum, which is defined predominantly by our choice of impact parameter, relative velocity at infinity and initial orbital separations, which we assume are all fixed.  We intend to explore this issue in more detail in future work.  Naively, we expect our basic formalism to still apply for larger encounter angular momenta, provided ergodicity is upheld.  We caution, however, that in the limit of very large differences between the initial binary orbital separations, the characteristic behaviours identified here could change.  In particular, previous work has shown that in this limit, the compact binary is better approximated as a single object, and the interaction is better described by the three-body problem \citep[e.g.][]{valtonen06,leigh16c}.  More work is needed to better understand these interesting details of the disruption of chaotic four-body systems.

\subsection{Astrophysical motivation} \label{astro}

In this section, we address possible applications of our results to outstanding problems in modern astrophysics.

\subsubsection{Star clusters} \label{clusters}

Binary-binary interactions occur commonly in dense star clusters, with binary fractions $f_{\rm b} \gtrsim 10\%$ \citep{leigh11,geller15}.  Given that a distribution of binary orbital parameters is expected for the binary population in a given star cluster, a range of initial combinations of binary orbital separations, eccentricities, mass ratios, etc. is expected.  For our numerical scattering experiments, we have chosen the initial velocity at infinity to be representative of what is actually expected in typical star cluster environments.  However, our results should be extended to actual star cluster environments with caution, since we explore a limited set of initial conditions in this paper which do not typically reflect a realistic star cluster environment (e.g., zero impact parameter).  Instead, our results represent only a subset of the types of binary-binary interactions expected in real star cluster environments, ranging from open clusters to dense globular clusters.  

\subsubsection{Black hole dynamics} \label{BHs}

One especially timely application of our results would be to populations of black holes (BHs) in dense star clusters, which are thought to act as dynamical factories for BH-BH binary mergers.  Recently, researchers of the aLIGO project have detected the presence of gravitational waves emitted from binary stellar-mass black hole coalescence \citep{abbott16}.  In the early 1990s, \citet{kulkarni93} and others argued that in globular clusters especially, mass segregation can in some extreme cases mediate a strong \lq slingshot" effect between heavy black holes interacting in the core of the cluster, which can eject the BHs from the cluster.  Though in this scenario most of the black hole population is depleted due to these effects, those that do remain (and some that are ejected) are locked in close-binary orbits.  More recent studies have amended these earlier works to show that significant BH systems can persist in GCs up to the present-day, particularly in the most massive lowest density GCs with the longest half-mass relaxation times. These are ideal candidates for the production of gravitational waves detectable by aLIGO \citep[e.g.][]{abbott16,antonini16,askar17}.  Rather fortuitously, \citet{arcasedda17} showed that the observable properties of the host GC can be used to infer the properties of a given GC's BH population.

The above picture relates directly to our results, which have immediate implications for exchange interactions producing BH-BH binaries in globular clusters.  We have shown that there is a direct relation between the lifetime distribution for a given outcome and the probability of obtaining that outcome in the final state.  The encounter duration is an important parameter to consider in globular clusters where the central density is sufficiently high that direct encounters involving binaries and single stars occur commonly, perhaps mediating the direct interruption of prolonged encounters by intervening single and binary stars \citep[e.g.][]{geller15,leigh16a}.  

More generally, the half-life formalism presented here for the chaotic four-body problem can, in principle, be extended to larger-$N$ systems of particles.  The ideal astrophysical application for this method are mass segregated populations of stellar-mass black holes in dense globular and nuclear clusters.  These BHs represent heavy particles that have typically decoupled dynamically from the lighter background stars, forming their own bound sub-system of particles in the cluster core.  As already described, these BHs then undergo strong encounters and \lq\lq "slingshot" each other to large velocities, becoming unbound and escaping from their host cluster.  In principle, each of these ejection events can be regarded as a decay, with a corresponding half-life. This is because, just as radioactive isotopes decay into more stable daughter products, so does an $N$-body system decay (ultimately) into a stable set of gravitationally-bound systems. If the half-lives depend only on the numbers and masses of each type of BH, then it might be possible to use a half-life formalism to predict the probability for a given sequence of ejection events. More generally, by understanding the distribution of ejection or disruption times, physical insight can be gained for application to the disruption time distributions of larger-$N$ systems.  The dependence of these distributions on the properties of the BH sub-system and the host GC can then be looked for.  We intend to further consider this interesting possibility in a future paper.

\subsection{Understanding the underlying physical mechanism} \label{generic}

Our results suggest that there is a relation between the lifetime distribution corresponding to a particular outcome of a decaying four-body system and the probability of the interaction ending in that outcome.  What is the physical reason for this correlation?  We speculate on a few possible reasons here.  First, consider a scenario in which, from time-step to time-step, the four-body system jumps around in phase space with a uniform probability of ending up anywhere in this phase space between successive time-steps.  In this case, the probability of the interaction ending in a given outcome is directly proportional to the volume of phase space corresponding to that outcome.  This is one scenario that would be consistent with our results, but this does not necessarily mean it is the correct mechanism.  

Now consider a different scenario in which the system evolves in time through phase space by performing a random walk from its initial location in this phase space.  In this case, the relation between the lifetime of the system and the probability of obtaining a given outcome is more complicated but, similar to before, we would naively expect a positive correlation between the probability of a given outcome and the volume in phase space corresponding to that outcome.  

The key point is that different behaviours for the time evolution of the system through phase space should naively be reflected in the subsequent disruption time distributions.  More work will be needed to better understand degenerate modes for the time evolution of the system through phase space yielding indistinguishable disruption time distributions.  We intend to study these issues further in future work, in an effort to better constrain the true physical mechanism(s) responsible for the interesting connection found in this paper between the lifetime distribution corresponding to a particular outcome of a decaying four-body system and the probability of the interaction ending in that outcome.

Interestingly, the above discussion could facilitate a more empirical test of the idea of using the disruption of gravitationally-bound $N$-body systems as a possible probe of the underlying physics governing the disruption of other $N$-body systems mediated by small-range forces operating on microscopic scales. For example, this could include radioactive decay and other disrupting systems obeying an exponential distribution of disruption times, as found for small-number self-gravitating systems \citep[e.g.][]{leigh16b}.  This is because a given distribution of disruption times directly constrains the evolution of the system through phase space.  We emphasize, however, that this avenue should be explored with caution, given that extensive degeneracies could exist (among other issues).  In order to explore this idea further, appropriate empirical tests need to be developed.  As described above, these tests could be based on hypotheses for the underlying manner by which the evolving system moves around in phase space, by re-distributing the energies and angular momenta of the interacting particles.  

\section{Summary} \label{summary}

In this Part II of the series, we consider the possibility of $N$-body systems sharing dynamical behavior reminiscent of radioactive decay, by studying the distribution of disruption times during binary-binary interactions. This is done using a half-life formalism, applied independently to each of the three possible outcomes.  We obtain estimates for the half-life of each outcome using numerical simulations and use these to calculate expectations for the relative outcome fractions, which we compare directly to the simulations. We consider different particle mass combinations, but fix the total encounter energy.  The ratio of the calculated half-lives obtained from Gaussian fits to the simulated lifetime distributions are consistent with the simulated branching ratios or relative outcome fractions, thus supporting our model and hypothesis.  We briefly discuss the application of our results to the evaporation of black hole sub-clusters in globular clusters.

\section*{Acknowledgments}

N.~W.~C.~L. acknowledges support from a Kalbfleisch Fellowship at the American Museum of Natural History and the Richard Gilder Graduate School,  as well as support from National Science Foundation Award AST 11-09395.  
T.~I. is grateful to Dr. John Davis for his guidance and salutary suggestions that helped make this effort possible.

\appendix
\section{Plots for {\bfseries $m_1=3\Msol$ and $5\Msol$}}
\label{Appendix}

\begin{figure}
\centering\includegraphics[width=\columnwidth]{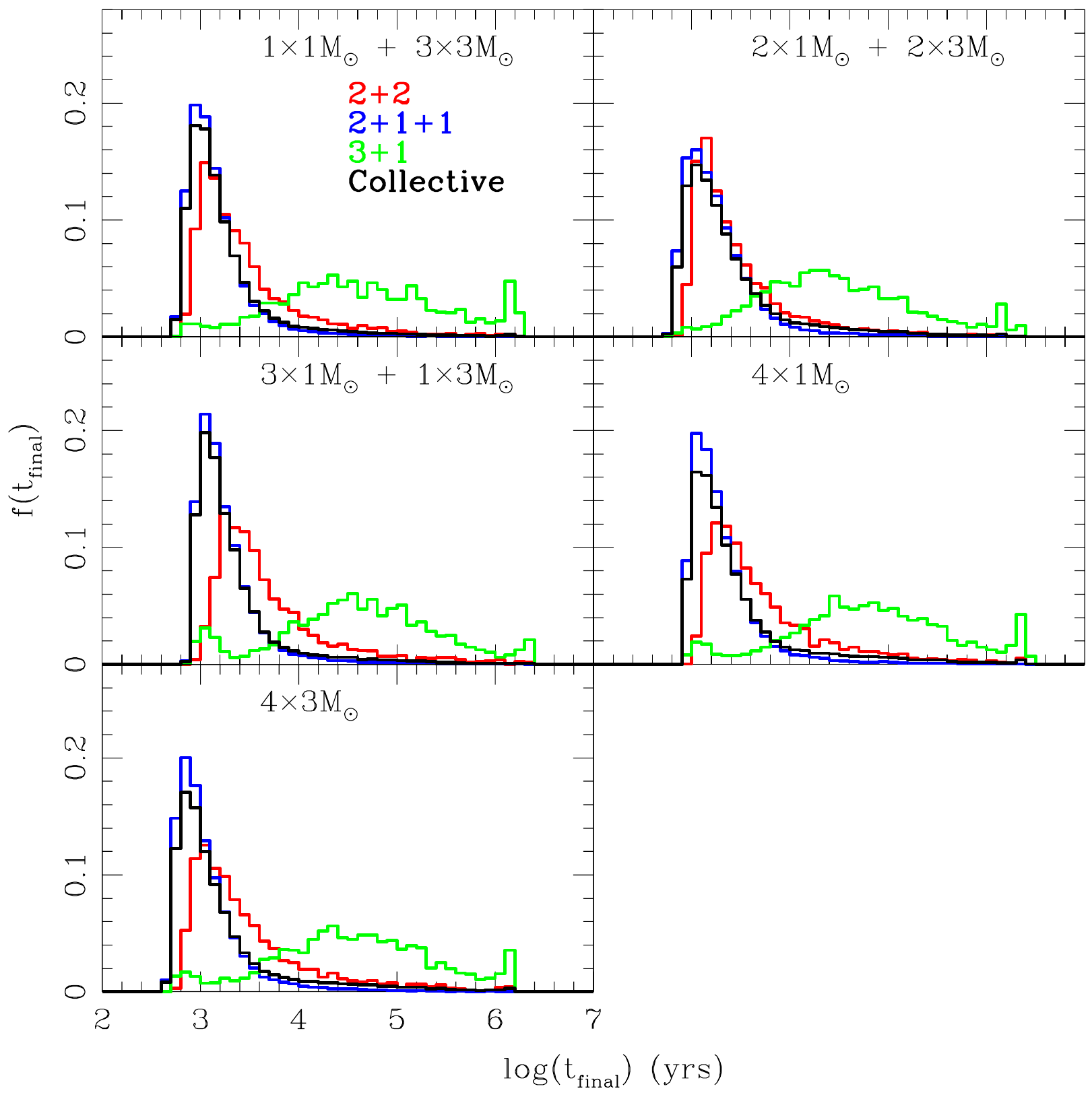}
\caption[Combined distributions before partitioning into resultants for the mass ratio $m_1/m_2=3$]{Analog to Figure \ref{fig:10CHistograms}, except that $m_1=3\Msol$.}
\label{fig:3CHistograms}
\end{figure}

\begin{figure}
\centering\includegraphics[width=\columnwidth]{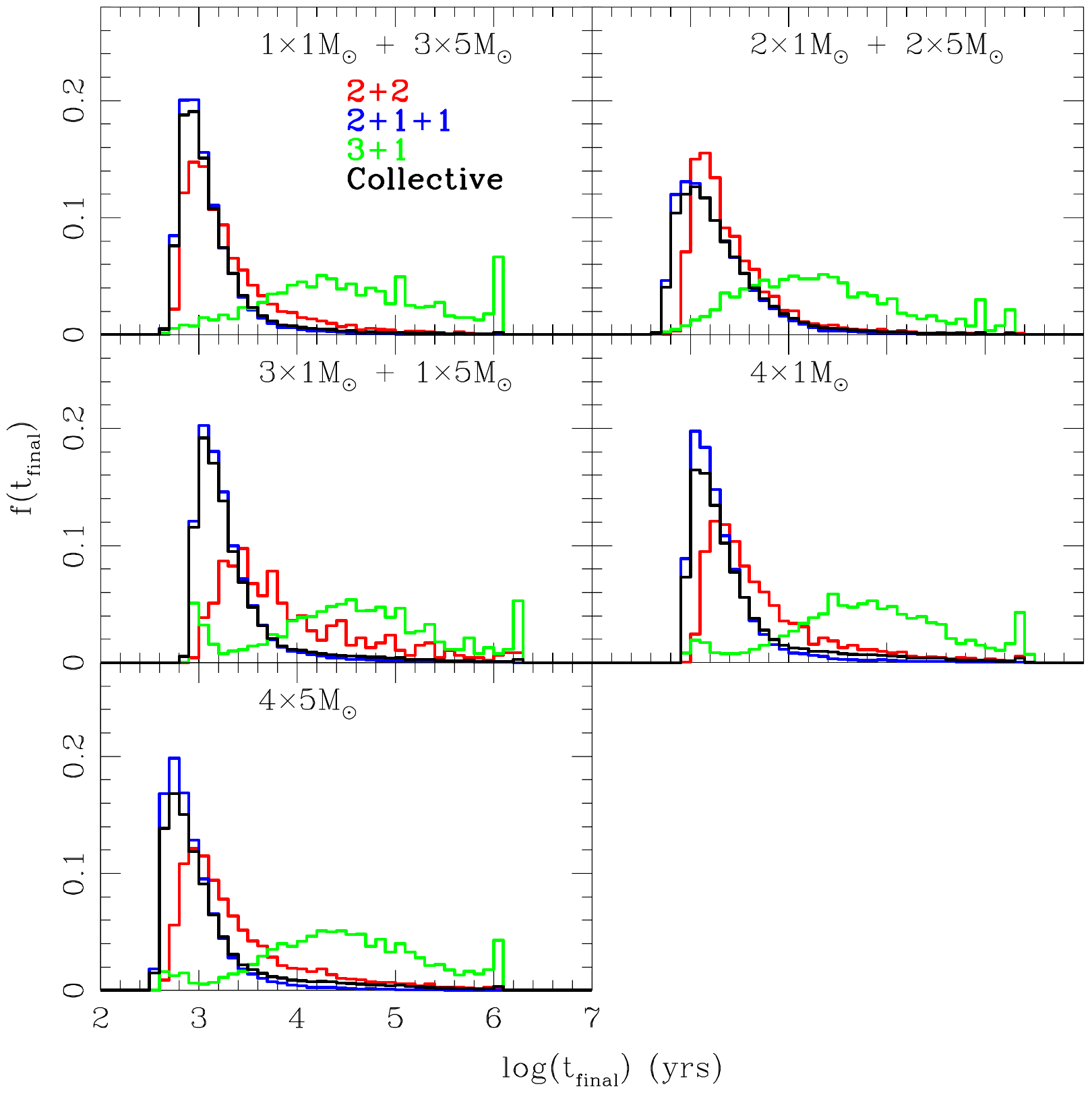}
\caption[Combined distributions before partitioning into resultants for the mass ratio $m_1/m_2=5$]{Analog to Figure \ref{fig:10CHistograms}, except that $m_1=5\Msol$.}
\label{fig:5CHistograms}
\end{figure}

\begin{figure}
\centering\includegraphics[width=\columnwidth]{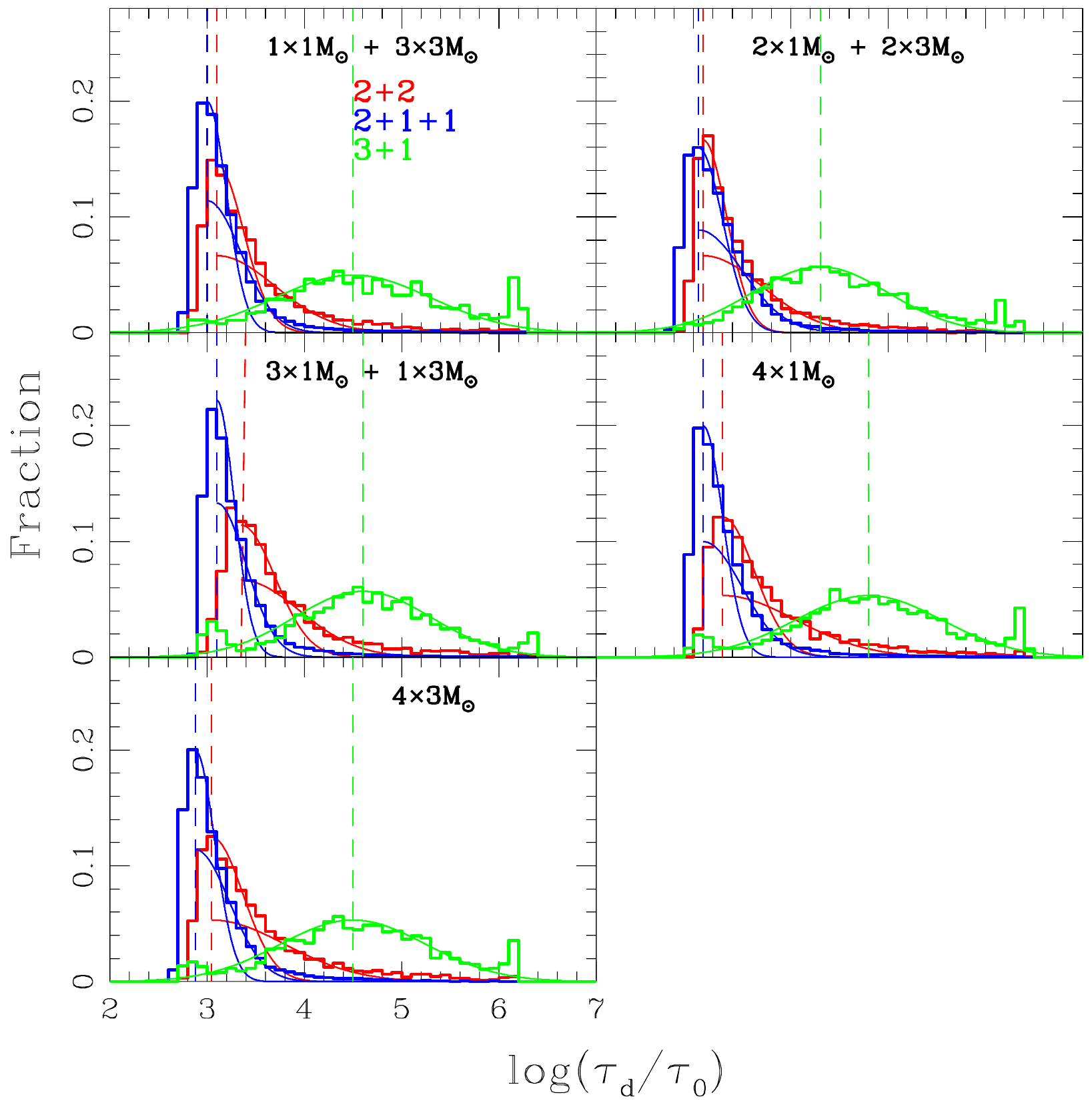}
\caption[The distributions of encounter durations for each possible outcome state, and for all combinations of light and heavy particles, with $m_{\rm 1} =3\Msol$]{Analog to Figure~\ref{fig:10Models} except with $m_1=3\Msol$}
\label{fig:3Models}
\end{figure}

\begin{figure}
\centering\includegraphics[width=\columnwidth]{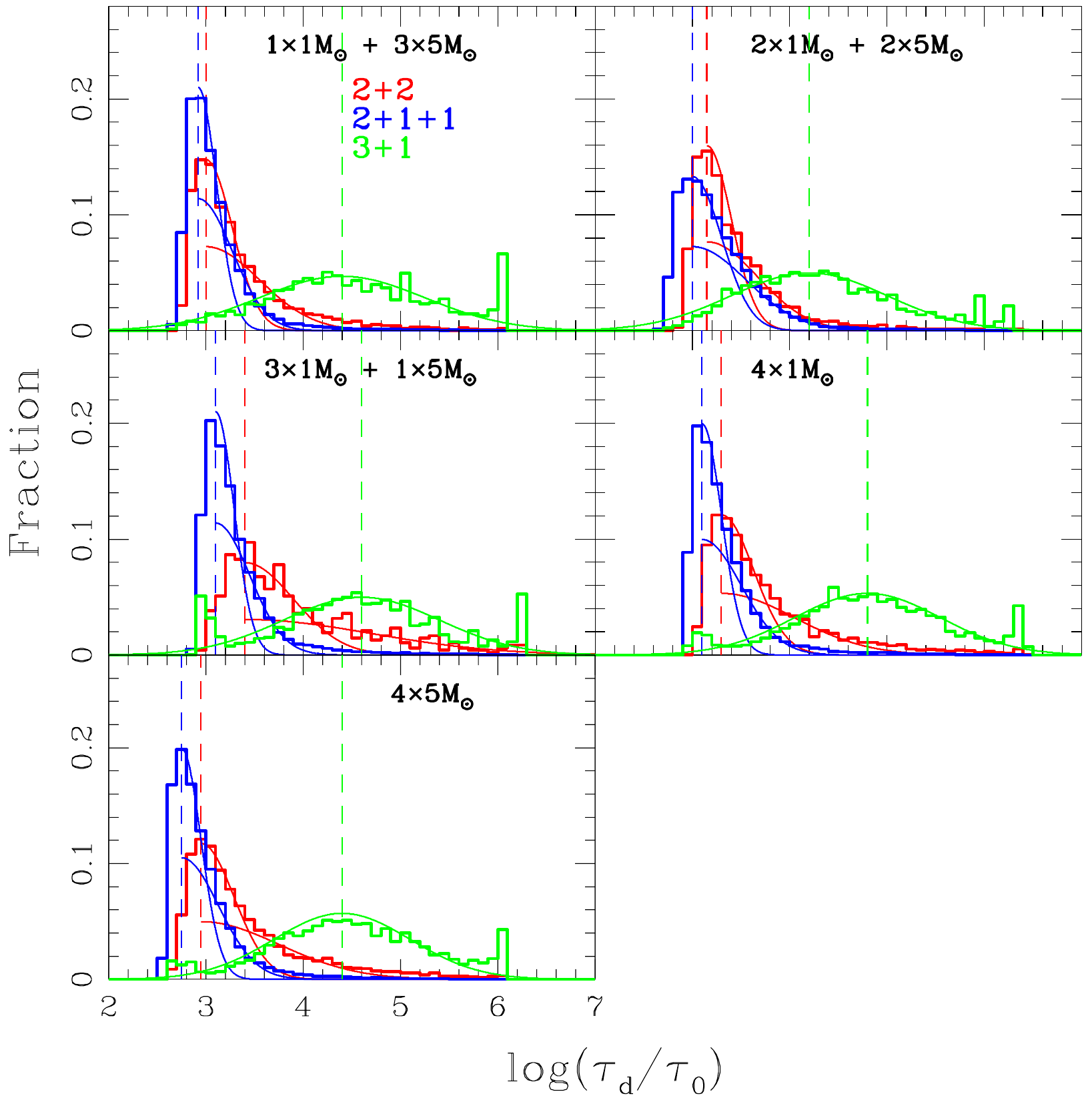}
\caption[The distributions of encounter durations for each possible outcome state, and for all combinations of light and heavy particles, with $m_{\rm 1} = 5 \Msol$]{Analog to Figure~\ref{fig:10Models} except with $m_1=5\Msol$}
\label{fig:5Models}
\end{figure}

\begin{figure}
\begin{center}                                                                                                                                                           
\includegraphics[width=\columnwidth]{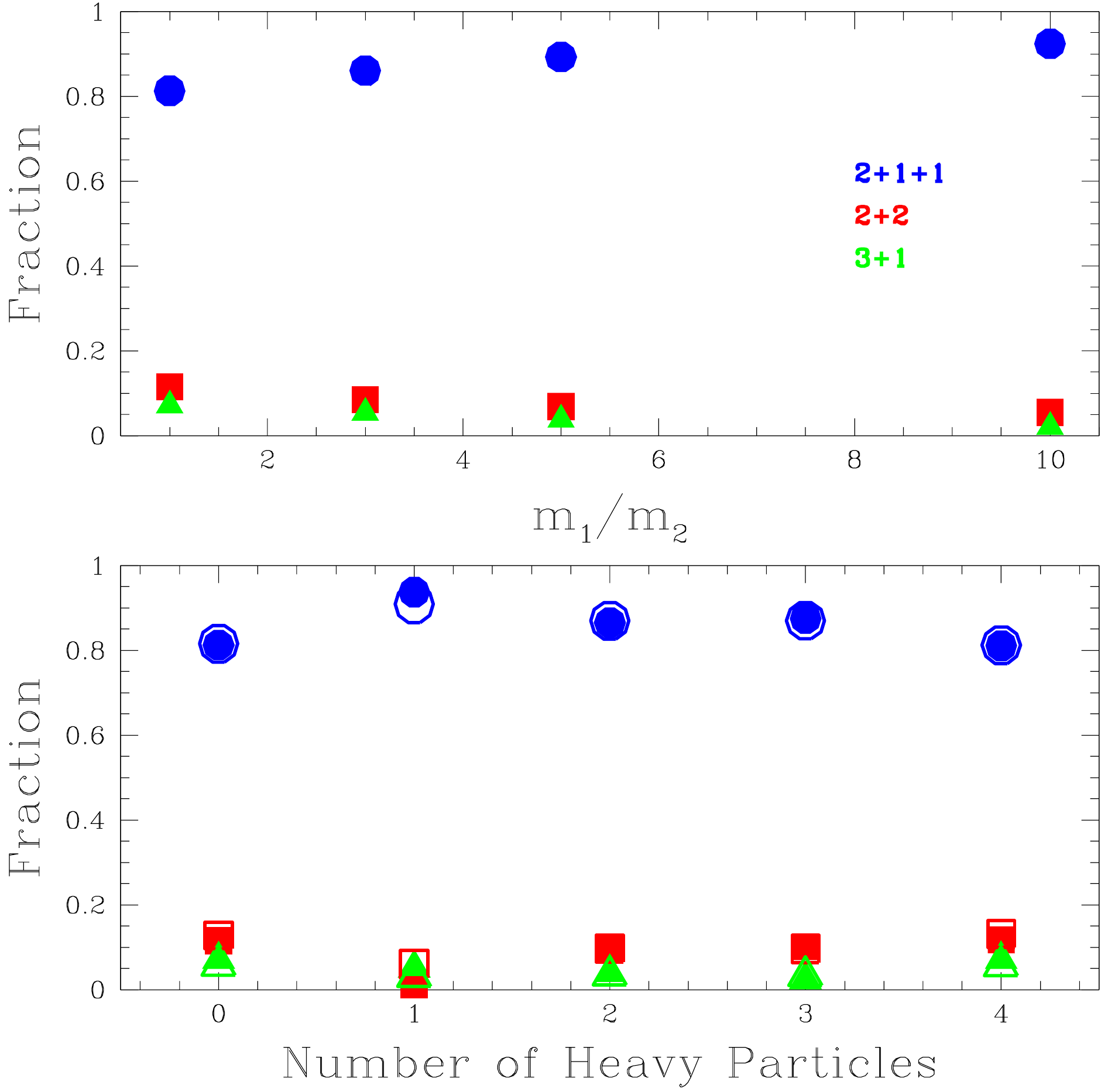}
\end{center}
\caption[The fractions of binary-binary interactions that produce 3+1, 2+2 and 2+1+1 outcomes, as a function of the distributions of particle masses, with $m_1=3\Msol$]{Analog to Figure \ref{fig:10Fractions}, except for $m_1=3\Msol$. Again notice the near-perfect agreement of the branching ratios from our hypothesized half-lives to the simulated ratios.}
\label{fig:3Fractions}
\end{figure}

\begin{figure}
\begin{center}                                                                                                                                                           
\includegraphics[width=\columnwidth]{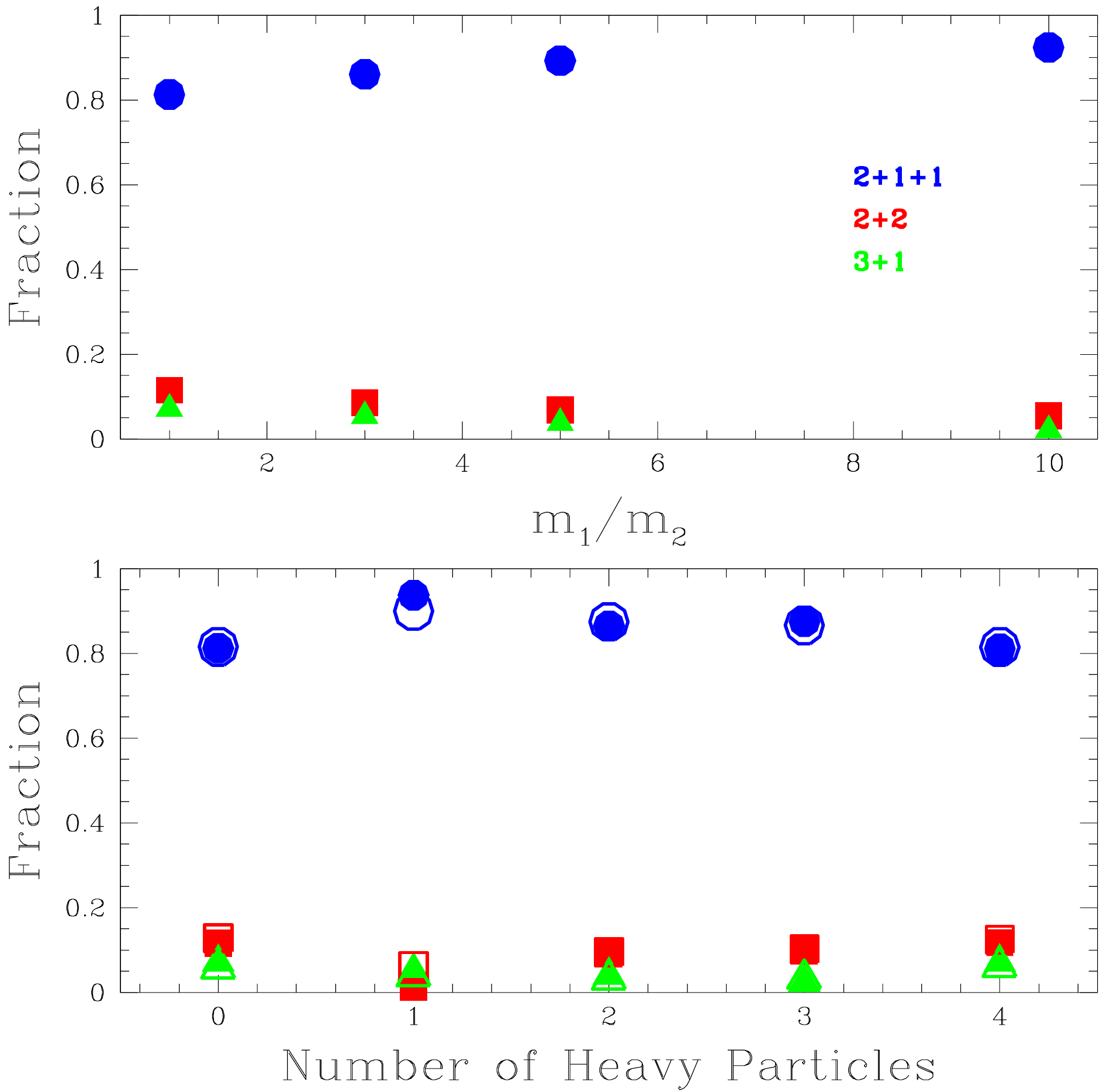}
\end{center}
\caption[The fractions of binary-binary interactions that produce 3+1, 2+2 and 2+1+1 outcomes, as a function of the distributions of particle masses, with $m_1=5\Msol$]{Analog to Figure~\ref{fig:10Fractions}, except for $m_1=5\Msol$. Just as in the previous plots, the agreement of the superimposed open figures corroborates our model established in \textsection~\ref{virial}.}
\label{fig:5Fractions}
\end{figure}

\begin{figure}
\begin{center}
\includegraphics[width=\columnwidth]{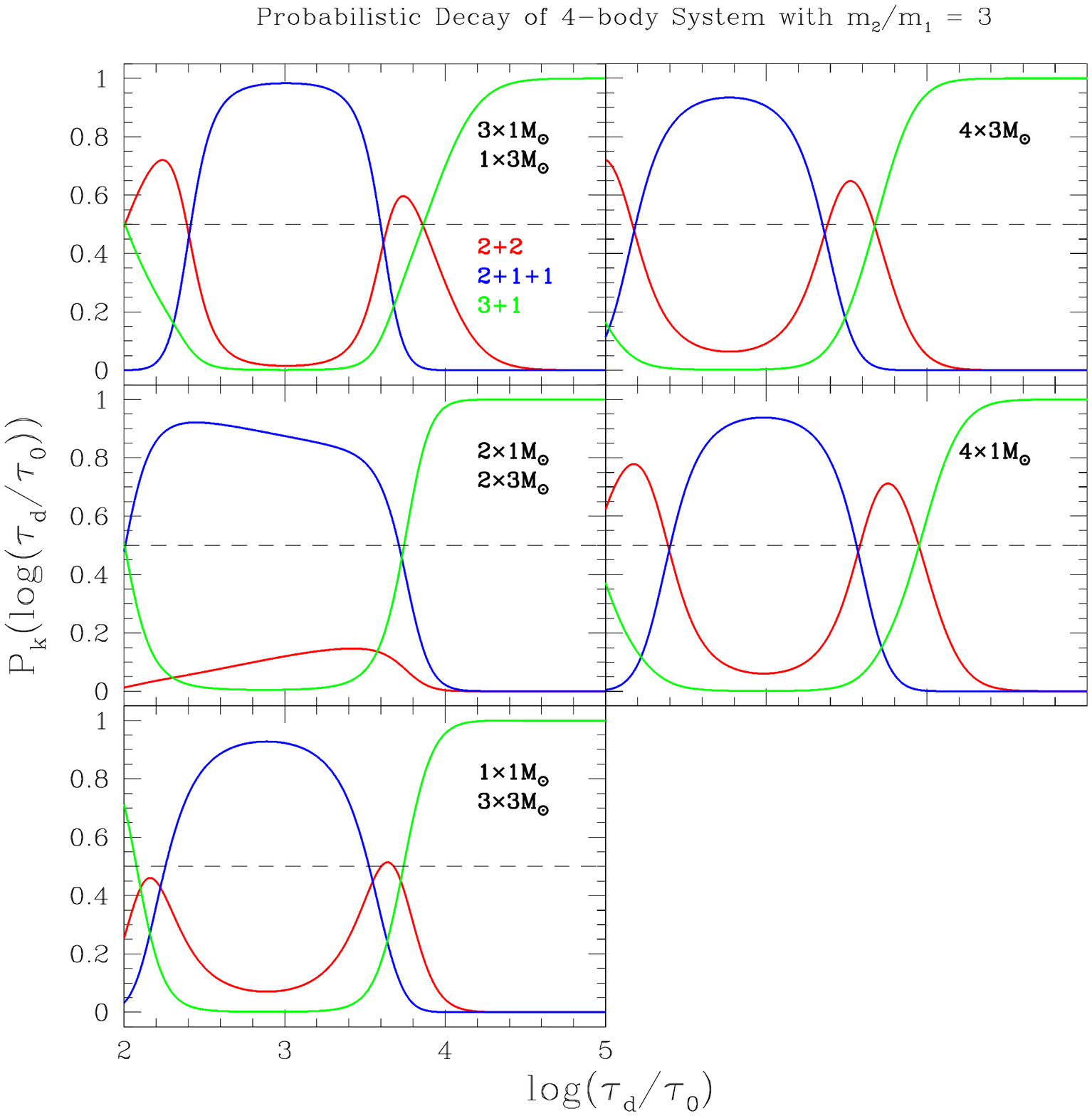}
\end{center}
\caption{Analog to Figure \ref{fig:10Probabilistic}, save for that $m_1=3\Msol$.}
\label{fig:3Probabilistic}
\end{figure}

\begin{figure}
\begin{center}
\includegraphics[width=\columnwidth]{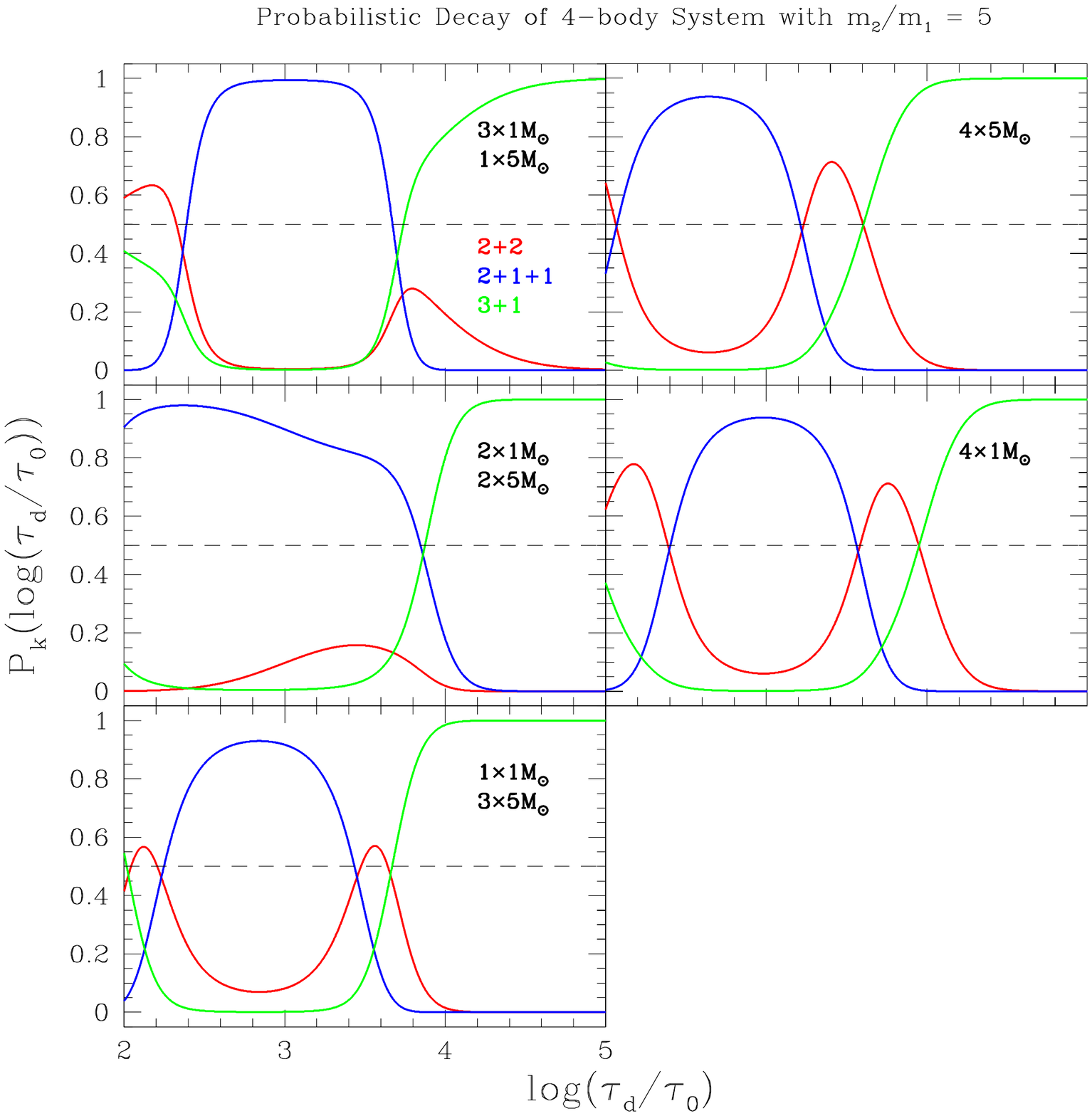}
\end{center}
\caption{Analog to Figure~\ref{fig:10Probabilistic}, save for that $m_1=5\Msol$. Notice that the plots with an even distribution of each mass in a mass combination yield the most distinct patterns of probabilities, while the other plots follow a general trend of alteration.}
\label{fig:5Probabilistic}
\end{figure}

\bsp   

\label{lastpage}

\end{document}